\documentclass{emulateapj}

\begin{document}
   \title{The Herschel Reference Survey}


   \author{A. Boselli\altaffilmark{1}
          ,
          S. Eales\altaffilmark{2}
	  ,
	  L. Cortese\altaffilmark{2}
	  , 
	  G. Bendo\altaffilmark{3}
	  ,	  
	  P. Chanial\altaffilmark{3}
	  ,
	  V. Buat\altaffilmark{1}
	  ,
	  J. Davies\altaffilmark{2}
	  ,
	  R. Auld\altaffilmark{2}
	  ,
	  E. Rigby\altaffilmark{4}
	  ,
	  M. Baes\altaffilmark{5}
	  ,
	  M. Barlow\altaffilmark{6}
	  ,
	  J. Bock\altaffilmark{7}
	  ,
	  M. Bradford\altaffilmark{7}
	  ,
	  N. Castro-Rodriguez\altaffilmark{8}
	  ,
	  S. Charlot\altaffilmark{9}
	  ,
	  D. Clements\altaffilmark{3}
	  ,
	  D. Cormier\altaffilmark{11}
	  ,
	  E. Dwek\altaffilmark{10}
	  ,
	  D. Elbaz\altaffilmark{11}
	  ,
	  M. Galametz\altaffilmark{11}
	  ,
	  F. Galliano\altaffilmark{12}
	  ,
	  W. Gear\altaffilmark{2}
	  ,
	  J. Glenn\altaffilmark{13}
	  ,
	  H. Gomez\altaffilmark{2}
	  ,
	  M. Griffin\altaffilmark{2}
	  ,
	  S. Hony\altaffilmark{11}
	  ,
	  K. Isaak\altaffilmark{2}
	  ,
	  L. Levenson\altaffilmark{7}
	   ,
	  N. Lu\altaffilmark{7}
	  ,
	  S. Madden\altaffilmark{11}
	  ,
	  B. O'Halloran\altaffilmark{3}
	  ,
	  K. Okumura\altaffilmark{11}
	  ,
	  S. Oliver\altaffilmark{14}
	  ,
	  M. Page\altaffilmark{15}
	  ,
	  P. Panuzzo\altaffilmark{11}
	  ,
	  A. Papageorgiou\altaffilmark{2}
	  ,
	  T. Parkin\altaffilmark{20}
	  ,
	  I. Perez-Fournon\altaffilmark{8}
	  ,
	  M. Pohlen\altaffilmark{2}
	  ,
	  N. Rangwala\altaffilmark{13}
	  ,
	  H. Roussel\altaffilmark{9}
	  ,
	  A. Rykala\altaffilmark{2}
	  ,
	  N. Sacchi\altaffilmark{17}
	  ,
	  M. Sauvage\altaffilmark{11}
	  ,
	  B. Schulz\altaffilmark{16}
	  ,
	  M. Schirm\altaffilmark{20}
	  ,
	  M.W.L. Smith\altaffilmark{2}
	  ,
	  L. Spinoglio\altaffilmark{17}
	  ,
	  J. Stevens\altaffilmark{18}
	  ,
	  M. Symeonidis\altaffilmark{18}
	  ,
	  M. Vaccari\altaffilmark{19}
	  ,
	  L. Vigroux\altaffilmark{9}
	  ,
	  C. Wilson\altaffilmark{20}
	  ,
	  H. Wozniak\altaffilmark{21}
	  ,
	  G. Wright\altaffilmark{18}
	  ,
	  W. Zeilinger\altaffilmark{22}
         }
\altaffiltext{1}{Laboratoire d'Astrophysique de Marseille, UMR6110 CNRS, 38 rue F. Joliot-Curie, F-13388 Marseille France}
\altaffiltext{2}{School of Physics and Astronomy, Cardiff University, Queens Buildings The Parade, Cardiff CF24 3AA, UK}
\altaffiltext{3}{Astrophysics Group, Imperial College, Blackett Laboratory, Prince Consort Road, London SW7 2AZ, UK}
\altaffiltext{4}{School of Physics \& Astronomy, University of Nottingham, University Park, Nottingham NG7 2RD, UK}
\altaffiltext{5}{Sterrenkundig Observatorium, Universiteit Gent, Krijgslaan 281 S9, B-9000 Gent, Belgium}
\altaffiltext{6}{Department of Physics and Astronomy, University College London, Gower Street, London WC1E 6BT, UK}
\altaffiltext{7}{Jet Propulsion Laboratory, Pasadena, CA 91109, United States; Department of Astronomy, California Institute of Technology, Pasadena, CA 91125, USA}
\altaffiltext{8}{Instituto de Astrofísica de Canarias, C/Vía Láctea s/n, E-38200 La Laguna, Spain}
\altaffiltext{9}{Institut d'Astrophysique de Paris, UMR7095 CNRS, Universit\'e Pierre \& Marie Curie, 98 bis Boulevard Arago, F-75014 Paris, France}
\altaffiltext{10}{Observational  Cosmology Lab, Code 665, NASA Goddard Space Flight  Center Greenbelt, MD 20771, USA}
\altaffiltext{11}{Laboratoire AIM, CEA/DSM - CNRS - Universit\'e Paris Diderot, Irfu/Service d'Astrophysique, 91191 Gif sur Yvette, France}
\altaffiltext{12}{Department of Astronomy, University of Maryland, College Park, MD20742,  USA}
\altaffiltext{13}{Department of Astrophysical and Planetary Sciences, CASA CB-389, University of Colorado, Boulder, CO 80309, USA}
\altaffiltext{14}{Astronomy Centre, Department of Physics and Astronomy, University of Sussex, UK}
\altaffiltext{15}{Mullard Space Science Laboratory, University College London, Holmbury St Mary, Dorking, Surrey RH5 6NT, UK}
\altaffiltext{16}{Infrared Processing and Analysis Center, California Institute of Technology, Mail Code 100-22, 770 South Wilson Av, Pasadena, CA 91125, USA}
\altaffiltext{17}{Istituto di Fisica dello Spazio Interplanetario, INAF, Via del Fosso del Cavaliere 100, I-00133 Roma, Italy}
\altaffiltext{18}{Centre for Astrophysics Research, Science and Technology Research Centre, University of Hertfordshire, College Lane, Herts AL10 9AB, UK}
\altaffiltext{19}{University of Padova, Department of Astronomy, Vicolo Osservatorio 3, I-35122 Padova, Italy}
\altaffiltext{20}{Dept. of Physics \& Astronomy, McMaster University, Hamilton, Ontario, L8S 4M1, Canada}
\altaffiltext{21}{Observatoire Astronomique de Strasbourg, UMR 7550 Université de Strasbourg - CNRS, 11, rue de l'Universit\'e, F-67000 Strasbourg}
\altaffiltext{22}{Institut für Astronomie, Universität Wien, Türkenschanzstr. 17, A-1180 Wien, Austria}

\begin{abstract}

The Herschel Reference Survey is a guaranteed time Herschel key project and will be a benchmark study of dust in the nearby 
universe. The survey will complement a number of other Herschel key projects including large cosmological surveys that trace dust in the distant universe. 
We will use Herschel to produce images of a statistically-complete sample of 323 galaxies at 250, 350 and
500 $\mu$m. The sample is volume-limited, containing sources with distances between 15 and 25 Mpc and flux limits in the K-band
to minimize the selection effects
associated with dust and with young high-mass stars and to introduce a selection in stellar mass. 
The sample spans the whole range of morphological types (ellipticals to
late-type spirals) and environments (from the field to the centre of the Virgo Cluster) and as such will be useful 
for other purposes than our own. We plan to use the survey to investigate (i) the dust content of galaxies as a function
of Hubble type, stellar mass and environment, (ii) the connection between the dust content and composition and the other phases of the
interstellar medium and (iii) the origin and evolution of dust in galaxies. In this paper, we describe the goals of the
survey, the details of the sample and some of the auxiliary observing programs that we have started to collect
complementary data. We also use the available multi-frequency data to carry out an analysis of the statistical properties
of the sample.
\end{abstract}


\setcounter{footnote}{0}

\section{Introduction}

Understanding the processes that govern the formation and evolution of 
galaxies is one of the major challenges of modern astronomy. 
Ideally this work can be done by analyzing and comparing the 
physical, structural and kinematical properties of different objects at 
various epochs to model predictions. How primordial galaxies transform their huge gas reservoir into stars and 
become the objects that we now observe in the local universe is a key question.
A wide coverage of the electromagnetic spectrum 
to probe the atomic (21 cm line) and molecular (generally through the 2.6 mm CO line) 
gas components, 
the different stellar populations (UV to near-IR spectral range), 
the dust (mid- and far-IR) and the electrons in magnetic fields (radio continuum)
is necessary to address this important question.\\

The importance of dust resides in the fact that it 
is formed from the aggregation of the metals produced in the latest phases 
of stellar evolution, and contains approximately half the metals in the interstellar
medium (Whittet 1992). Injected into the interstellar
medium (ISM) by stellar winds and supernovae explosions, dust acts as a
catalyst in the process of transformation of the atomic to molecular hydrogen
necessary to feed star formation (Hollenbach \& Salpeter 1971; Duley \& Williams 1986). Dust also contributes to the shielding 
of the UV radiation field, preventing the dissociation of
molecular clouds, and thus playing a major role in the energetic equilibrium
of the ISM (Hollenbach \& Tielens 1997). Dust contributes to the cooling and heating of the
ISM in photodissociation regions through photoelectric heating. 
Furthermore, by absorbing the stellar light, dust modifies our
view of the different stellar components (e.g. Buat \& Xu 1996): 
because dust obscuration is so important
in star-forming regions, the emission from dust is one of the
most powerful tracers of the star formation activity in galaxies 
(Kennicutt 1998; Hirashita et al. 2003). Dust emission can therefore be used
to study the relationship between the gas surface density and the star formation
activity, generally known as the Schmidt law.\\

The importance of dust in the study of the formation and evolution of 
galaxies became evident after the IRAS space mission which provided us with
a whole sky coverage in four infrared bands. Despite its low sensitivity and poor
angular resolution, IRAS detected tens thousands of extragalactic 
sources (Soifer et al. 1987). The IRAS survey showed that the stellar light in
some galaxies is so heavily obscured that the only way to determine their star formation activity 
is through the dust emission itself. Among these highly obscured galaxies, the Ultraluminous Infrared Galaxies (ULIRGs) are
the most actively star-forming objects in the local universe. With their improved sensitivity,
spectral and angular resolution, other space missions
such as ISO and Spitzer have brought a better understanding of the chemical
and physical properties of interstellar dust in a wide range of different 
galactic and extragalactic sources. 
IRAS, ISO, Spitzer and the recently launched AKARI 
satellite, however, are sensitive to dust emitting in the mid- ($\sim$ 5 $\mu$m) 
to far- ($\leq$ 200 $\mu$m) infrared domain. 
The 5-70 $\mu$m spectral range 
corresponds to the emission of the hot dust generally associated 
to star formation, while at longer wavelengths, up to 1 mm, the contribution of cold
dust becomes more and more important (Draine et al. 2007). The emission of the ISM in the range 
3 $\mu$m $\la$ $\lambda$ $\la$ 15 $\mu$m is generally dominated by  
Polycyclic Aromatic Hydrocarbons (PAHs).
Between 10 $\mu$m and $\la$ 70 $\mu$m the dust emission 
is usually due to very small grains, while that at longer wavelengths
(70 $\mu$m $\la$ $\lambda$ $\la$ 1000 $\mu$m) is probably produced by big 
grains of graphite and silicate in thermal equilibrium 
with the UV and optical photons of the interstellar radiation field
(D\'esert et al. 1991; Dwek et al. 1997; Zubko et al. 2004; Draine \& Li 2007).\\

There is, however, strong empirical evidence to suggest that much of the dust in normal galaxies has been missed
by previous space missions because it is too
cold to radiate in the mid- and far-infrared. Devereux \& Young (1990), for example,
showed that when the dust masses of galaxies were estimated from IRAS data, the gas-to-dust ratios
were $\simeq$10 times higher than the Galactic value, implying that
90\% of the dust in galaxies is too cold to radiate significantly in the IRAS bands.
ISO 200 $\mu$m images of nearby galaxies revealed the presence of cold ($\sim$20 K)
dust in the external parts of galaxies (Alton et al. 1998).
Even Spitzer, with its longer wavelength coverage, is only sensitive to dust that is warmer than 
$\simeq$15 K (Bendo et al. 2003; Draine \& Li 2007; Draine et al. 2007). 
Simple estimates show that the temperature of a dust grain in thermal equilibrium in the
average interstellar radiation field is $\sim$ 20 K, which would produce a
modified black body spectrum with a peak at $\sim$ 200 $\mu$m with a flux rapidly decreasing
at longer wavelengths (Boselli et al. 2003; Dale et al. 2005, 2007; Draine \& Li 2007; Draine et al. 2007). It is 
important to remember, however, that most of the dust in a galaxy is likely to be at a much lower
temperature than 20 K because it is self-shielded, and so unaffected from the interstellar radiation field.
For example, in molecular clouds
the temperature of the dust is usually well below the canonical 20 K value once the 
average dust extinction ($A_V$) is greater than one, unless it is
close to a newly formed star (Ward-Thompson et al. 2002).
\\ 

The submillimetre waveband (200-1000 $\mu m$) is crucial for detecting this missing
cold dust component and thus for making accurate estimates of the total dust mass. 
Mass estimates made from far-infrared measurements are highly uncertain because of the
strong dependence of flux on dust temperature which is difficult to determine using data on the Wien 
side of the black body peak. At long wavelengths (Rayleigh-Jeans domain) the emissivity depends only on the
first power of the dust temperature, making it possible to use a submillimetre flux
to make an accurate estimate of the mass of dust (Eales et al. 1989;
Galliano et al. 2003; 2005).

Submillimetre continuum observations of galaxies have been made previously, in particular
with the ground-based submillimetre cameras, such as SCUBA on the James
Clerk Maxwell Telescope. 
These have confirmed that galaxies do indeed
contain a large amount of cold dust (Dunne \& Eales 2001; Galliano et al. 2003).
Until now, the largest submillimetre survey has been the
SCUBA Local Universe and Galaxy Survey (SLUGS), a survey of $\simeq$200 galaxies in two samples, one
selected from the IRAS survey and one selected at optical wavelengths. 
The survey produced the first estimates of the submillimetre luminosity and dust-mass (the space density
of galaxies as a function of dust mass) functions (Dunne et al. 2000; Vlahakis et al. 2005), 
and also indicated that some ellipticals are surprisingly strong submillimetre sources. 
Its limitation, and the limitation of all ground-based submillimetre
observations, is that of sensitivity: SLUGS struggled to detect galaxies not already detected
by IRAS.\\

The solution is to go into space and the Herschel space telescope is finally making this possible. 
In particular the relatively large field of view, high sensitivity, and coverage of a waveband
(250-500 $\mu m$) in which galaxies are intrinsically bright, make SPIRE 
on Herschel the ideal instrument for a study of all extragalactic sources 
(Griffin et al. 2006, 2007; Wasket et al. 2007). SPIRE has the sensitivity
to map large areas of the sky down to the confusion limit in quite modest
integration times, providing 
submillimetre data for large samples of galaxies at different redshifs, and is thus
well adapted for pointed observations for all kind of 
nearby extragalactic sources.\\

The SPIRE consortium has defined a number of coordinated guaranteed time programs
to get the best possible benefit of the unique Herschel
facilities for the study of galaxy evolution.
These include deep cosmological surveys, pointed observations of local galaxies
and surveys of representative regions in the Milky Way. To better characterize the
dust properties in the local universe the SPIRE extragalactic group has selected 
a volume-limited sample of
323 galaxies to be observed in guaranteed time in the three 
SPIRE bands at 250, 350 and 500 $\mu$m.
The importance of the local universe resides in the fact that 
it represents the endpoint of galaxy evolution, providing 
important boundary conditions to models and simulations.
Furthermore, at $\leq$ 30 Mpc the angular resolution
of SPIRE (a couple of kpc) allow us to resolve the different galaxy components
such as nuclei, bulges, discs and spiral arms.
Moreover, dwarf galaxies, by far the most common (yet still very poorly 
understood) galaxies in the universe, can only be observed locally.\\

A volume-limited sample was chosen as a way to limit distance dependent
selection biases. To limit any possible contamination due to the cosmic variance,
the volume should be much larger than the typical dimension of 
large-scale structures ($\sim$ 30 Mpc). At the same time 
it should be representative of the whole galaxy population inhabiting the local universe.
A near-infrared K-band selection was adopted
for two major reasons: (i) unlike optical surveys, which have strong selection effects due to the
presence of dust, near-IR data are only marginally affected by dust extinction; 
(ii) whereas the optical flux is highly dependent
on the number of relatively young stars and thus sensitive to recent episodes of star formation, 
the near-IR luminosity is a good measure
of the overall stellar mass (e.g. Gavazzi et al. 1996), which  recent studies suggest 
as the most important parameter in characterizing the statistical and evolutionary properties of galaxies. 
Indeed galaxy properties that appear to be tightly correlated with the galaxy's stellar mass 
include the following: physical properties, such as the
star formation activity, the gas content 
and the metallicity (Boselli et al. 2001; 2002; Zaritsky et al. 1994; 
Gavazzi et al. 2004; Tremonti et al. 2004); structural 
properties, such as the concentration index and the galaxy's light profile (Boselli et
al. 1997; Gavazzi et al. 2000; Scodeggio et al. 2002);
the amount and distribution of dark matter, as indicated by the Tully-Fisher relation and the shape
of the rotation curve for spirals (Tully \& Fisher 1977; Catinella et al. 2006) and
the fundamental plane for ellipticals (Dressler et al. 1987; 
Djorgovski \& Davis 1987); the stellar population, shown through the
colour-magnitude relations for early-type 
(Visvanathan \& Sandage 1977, Bower et al. 1992) and late-type galaxies
(Tully et al. 1982, Gavazzi et al. 1996) 
and the galaxy spectral energy distributions (Gavazzi et al. 2002, Kauffmann et al. 2003).
These correlations all indicate a down-sizing effect (e.g. Gavazzi et al. 1996; Cowie et al. 1996; Heavens et al.
2004), in which galaxies with high stellar masses formed most of their stars
at a much earlier cosmic epoch than those with low stellar masses. This underline the dominant role of mass, 
rather than morphology, in shaping galaxies.\\

The final sample includes galaxies with a large range
in luminosity (mass) and includes all Hubble types. Because of its definition, 
the selected sample also includes galaxies belonging to different 
density regions such as the core of the Virgo cluster, groups and pairs 
and isolated objects.
Given its completeness, the Herschel Reference Sample (HRS) will be a suitable 
reference for any statistical study. Combined with KINGFISH (the Herschel extension of SINGS, Kennicutt et al. 2003)
and VNGS (see below),
two samples optimized for the study of the different phases 
of the ISM in individual galaxies, these samples will provide the community with
a unique dataset for studying the ISM properties of galaxies in the local universe.\\

The paper is organized as follows. Section 2 describes  
the scientific goals of the survey. Section 3 gives a description of the 
selection criteria used to define the sample. Section 4 gives details of the
SPIRE observations we will carry out and Section 5 gives an overview of the multi-frequency data
that is available for the sample. Section 6 gives a brief description 
of the data processing and products.
The multifrequency statistical properties of the sample will be described in Section 7.

\section{Scientific objectives}

The overall goal of this Herschel survey is to improve our knowledge of the 
cold dust properties of the most common extragalactic sources populating the local universe.
Combined with multifrequency ancillary data covering the whole electromagnetic spectrum (see Sect. 5),
the new Herschel data will provide us with a unique dataset with which to:  \\

1) {\bf Trace, for the first time, the variation of the properties of the cold dust component (dust mass and temperature,
dust to gas ratio...) of normal galaxies along the Hubble sequence, and as a function of 
luminosity}. Given the large dispersion in galaxy properties (Morton \& Haynes 1994),
it is important the
sample to be large enough to contain representatives of all galaxy types and include both early- and late-type galaxies
spanning the largest possible range in luminosity. Where galaxies are resolved, we
can analyze the distribution of the cold dust in the different galaxy components, e.g. 
the nuclei, bulges, spiral arms and diffuse discs. This will provide important constraints on dust formation and evolution in galaxies
(Galliano et al. 2008).\\

2) {\bf Study the role of dust in the physics of the ISM}. As discussed in the introduction, dust plays a
major role in the energetic equilibrium of the ISM. It acts as catalyst for the formation of the 
molecular hydrogen (Hollenbach \& Salpeter 1971; Duley \& Williams 1986) and shields the molecular gas component preventing dissociation
from the diffuse interstellar radiation field (Hollenbach \& Tielens 1997). To understand the nature of the ISM we thus need to
know how the cold dust properties (temperature, composition, geometrical distribution...)
relate to other physical properties such as metallicity and interstellar radiation field (Boselli et al. 2004;
Galliano et al. 2008). 
By combining dust surface brightnesses with metallicity dependent gas to dust ratios and
HI column densities, sub-mm measurements can be used to determine H$_2$ column densities 
(Gu\'elin et al. 1993; 1995; Neininger et al. 1996; Boselli et al. 2002). SPIRE data will therefore provide us with an independent measure of the
molecular hydrogen component, and can be use for an accurate calibration of the still
highly uncertain CO to H$_2$ conversion factor. \\

3) {\bf Study the relationship between dust emission and star formation.}  
Dust participates in the
cooling of the gas through the shielding of the interstellar radiation field, in particular
of the UV light, and thus plays a major role in the process of star formation (Draine 1978; Dwek 1986; Hollenbach \& Tielens 1997).
The energy absorbed by dust is then radiated in the infrared domain.  
For the same reason dust emission has often been used as a tracer of star formation.
We still do not know, however, what is the relationship between the cold dust emission 
and star formation. The study of resolved galaxies will allow us to analyze the relationship
between the infrared surface brightness and the dust temperature (Chanial et al. 2007) and to trace
the connections between the star formation and the dust and gas properties at galactic scales, extending 
the recent results of Spitzer to all phases of the ISM (Gordon et al. 2004, 
Calzetti et al. 2005, 2007; Perez-Gonzalez et al. 2006; Kennicutt et al. 2007, Prescott et al. 2007; Thilker et al. 2007).\\

4) {\bf Study the dust extinction properties in galaxies}. Astronomers have tackled with the
problem of dust opacity in galaxies for over 40 years (e.g. Holmberg 1958; Disney et al. 1989;
Calzetti 2001), but have reached no consensus. The key problem is that optical
catalogs contain huge selection effects because of the existence of dust. We will be able to
address this issue in two ways. First, the submillimetre images will, for the first time, allow us to
determine the distribution of the dust column density in a very large number of galaxies.
Second, we will be able to determine the global dust opacity in each galaxy by using the energy
balance between the absorbed stellar light 
and the dust emitted radiation (Buat \& Xu 1996, Witt \& Gordon 2000, Buat et al. 2002, 
Cortese et al. 2006; 2008a). This requires an accurate determination of
the UV to near-IR (stellar light) and mid-IR to sub-mm (dust emission) spectral energy 
distribution (Boselli et al. 2003). By comparing the dust attenuation properties of different classes of objects 
this analysis will allow us to define standard recipes for correcting UV 
and optical data, a useful tool for
the interpretation of all modern surveys.\\

5) {\bf Determine whether there is an intergalactic dust cycle.} 
Apart from the obvious possibility that dust is
ejected from galaxies by the same methods that gas is ejected, such
as through interactions with the surrounding environment (see point 9) 
and starburst-driven 'superwinds' as indeed observed in M82 (Engelbracht et al. 2006), there is also the
possibility that dust is ejected from galaxies by radiation pressure
(Davies et al. 1998, Meiksin 2009, Oppenheimer \& Dav\'e 2008). 
The ejection of dust from galaxies might explain the huge
reservoir of metals found in the intergalactic medium (Rayan-Weber et al. 2006). There is 
clear evidence for dust in superwinds (Alton et al. 1999) and there is tantalizing evidence
from ISO observations for extended distributions of dust around galaxies (Alton et al.
1998; Davies et al. 1999). Observations so far have been limited by
either sensitivity (SCUBA) or resolution (ISO and Spitzer). 
The Herschel Reference Survey will be able to determine
whether dust ejection is important because (i) we will be observing several hundred galaxies, so even if
these phenomena are rare our sample should contain some examples, and (ii) our observations will
have the sensitivity to detect dust well outside the optical disc of each galaxy.\\

6) {\bf Determine the amount of interstellar dust in ellipticals.} 
Very little is known about dust in ellipticals. Despite the stereotype that ellipticals
do not contain any dust, the structures seen in optical images imply that at least 50\% of 
ellipticals contain some dust (Ferrarese et al. 2006). The amount of interstellar dust is too small, however,
to be easily detected through its far infrared emission; IRAS, for example, detected only about
15\% of ellipticals (Bregman et al. 1998). Although Spitzer has now detected many ellipticals,
these studies have mostly been of sources known a priori to contain some dust
(Kaneda et al. 2007; Temi et al. 2007; Panuzzo et al. 2007) or molecular gas (Young et al. 2008). 
The SLUGS did contain a small statistically-complete sample
of 11 ellipticals, although some of the six 850-$\mu$m detections may well have been of synchrotron
rather than dust emission (Vlahakis et al. 2005). Other ellipticals have been detected at 350 $\mu$m
by Leeuw et al. (2008). The HRS contains
65 early-type (E, S0 and S0a) galaxies and our observations will have sufficient
sensitivity to detect dust masses down to $\rm \sim 10^4\ M_{\odot}$. The Herschel Reference Survey will therefore
provide an unambiguous answer to the question of how much dust is in ellipticals.\\

7) {\bf Determine the origin of dust in ellipticals.} The three possible origins of the dust in ellipticals are
that (i) it has been produced in the atmospheres of the old evolved stars that dominate the light of
ellipticals today, (ii) its origin is external to the elliptical and is the result of a merger,
(iii) it is the relic of dust formed during the active star-forming phase
early in the history of the galaxy. 
The distribution of dust is an important clue to understand its true origin. 
The presence of the first mechanism is suggested by Spitzer mid-infrared spectra of Virgo 
ellipticals (Bressan et al. 2006) which show the silicate emission produced by dust in circumstellar 
envelopes of evolved stars. The origin of this feature (predicted by Bressan et al. 1998) is 
supported by the finding that the mid-infrared emission has the same profile as optical light (Xilouris et al. 2004) 
in many early-type galaxies. It is unclear, however, whether this is the predominant 
mechanism producing the interstellar dust in these galaxies. For example, a submillimetre map of the nearby elliptical
Centaurus A, has revealed a dusty disc, implying that the dust (and the galaxy itself) has been formed
as the result of a merger (Leeuw et al. 2002). Temi et al. (2007) also found little correlation between the
Spitzer 70 and 160 $\mu$m emission and optical starlight, which also suggests
the dust has an external origin.\\

We will use the Herschel Reference Survey results to investigate the origin of the dust by, first,
investigating the detailed morphology of the dust, in particular
how it compares with the stellar distribution, and, second, by investigating whether the mass of dust is correlated
with other global properties of the galaxy such as stellar mass.
It has often been argued that sputtering by the ubiquitous, hot X-ray emitting gas 
in early galaxies should destroy any dust formed more than 10$^8$  years in the past (Tielens et al. 1994). 
However, the fact that dust is seen visually in 50\% of ellipticals (Ferrarese et al. 2006) and the 
recent discovery of small grains, which are preferentially destroyed by sputtering, in ellipticals 
(Kaneda et al. 2007) suggests that dust is protected in some way. We will investigate whether sputtering 
from the hot gas destroys the dust grains and will test the internal origin hypothesis by investigating 
any possible correlation between X-ray excess and dust mass.\\

8) {\bf Measure the local luminosity and dust-mass distributions.} These distributions are
important benchmarks for the deep Herschel surveys.
These have already been estimated as part of SLUGS (Dunne et al. 2000; Vlahakis et al. 2005): the
much greater sensitivity of the Herschel Reference Survey will push these limits down by a factor of $\sim$ 50.\\

9) {\bf Understand the role of the environment on the evolution of galaxies. }
The well-known phenomenon that spiral galaxies in clusters are HI-deficient and have 
truncated HI discs (eg. Cayatte et al. 1990) demonstrates that the 
environment of a galaxy can have a strong effect on its interstellar medium.
Indeed there is clear evidence for tidally-stripped dust in interacting galaxies (Thomas et al. 2002)
and indications for ram-pressure stripped dust in cluster objects (Boselli \& Gavazzi 2006).
At the same time the presence of metals (Sarazin 1986) and possibly 
of dust (Stickel et al. 2002; Montier \& Giard 2005)
in the diffuse intracluster medium has been shown by
X-ray and far-IR observations.
The comparison of the submillimetre emission of cluster and isolated galaxies within the HRS 
will allow us to make a detailed study on the effects of the
environment on the dust properties of galaxies, and thus understand whether the 
hot and dense cluster
intergalactic medium can be polluted through the gas stripping process of 
late-type galaxies (Boselli \& Gavazzi 2006).\\

10) {\bf Provide a multi-frequency reference sample suitable for any statistical study. }
Our aim is that the HRS will be the first large sample of galaxies with observations of all phases of the ISM, 
as well as measurements over the entire wavelength range of the
spectral energy distribution (SED) for each galaxy. The sample will then serve for
many purposes e.g. the  
UV to radio continuum SEDs could, for example, be used to determine the nature of unresolved  
sources or as templates for estimating photometric redshifts.\\

\section{The sample}

The Herschel Reference Sample (HRS) has been selected according to the following criteria:\\

1) {\bf Volume-limited}: A volume limit was imposed to reduce distance 
uncertainties due to local peculiar motions and ensure the presence of
low-luminosity, dwarf galaxies, not accessible at high redshift. By applying a lower distance ($D$)
limit we exclude the very extended sources, the observing of which
would be too time consuming \footnote{A small 
sample of very nearby and extended galaxies will be observed in detail as part of another 
guaranteed time project: Physical Processes in the Interstellar Medium
of Very Nearby Galaxies.}.  We have selected 
all galaxies with an optical recessional velocity ($vel$, taken from NED) in between 1050 km s$^{-1}$ and 1750 km s$^{-1}$ that, for 
$H_0$ = 70 km s$^{-1}$ Mpc$^{-1}$ and in the absence of peculiar motions,
corresponds to 15 $\leq$ $D$ $\leq$ 25 Mpc \footnote{Given the possible discrepancy between optical and HI
recessional velocity measurements, heliocentric velocities given in Table 1 can be outside this range.}. 
In the Virgo cluster region (12h$<$ R.A.(2000) $<$ 13h; 0$^\circ$$<$ dec $<$ 18$^\circ$), 
where peculiar motions are dominant, we have included all galaxies with
$vel$ $<$ 3000 km s$^{-1}$ and belonging to 
cluster A, the North (N) and East (E) clouds and the Southern extension (S) (17 Mpc)
and Virgo B (23 Mpc), where the subgroup membership has been taken from 
Gavazzi et al. (1999a). W and M clouds objects, at a distance of 32 Mpc, have been excluded 
\footnote{The sample was selected before distances were available on NED. According to these new NED estimates, 
the distance of the HRS galaxies outside the Virgo cluster are generally included in the 15 $\leq$ $D$ $\leq$ 25 Mpc
range. In the Virgo region, however, our distances are determined according to subgroup membership criteria,
and are thus generally different than those given by NED}. \\

2) {\bf K band selection}:
Given the expected low dust content of quiescent 
galaxies, whose emission would be hardly detectable within reasonable
integration times, a more stringent limit has been adopted for early-types  
than for star forming galaxies.
Among those galaxies in the required recessional velocity range, we thus selected  
all late-type spirals and irregulars (Sa-Sd-Im-BCD) with a 2MASS K band total magnitude $K_S$$_{tot}$ $\leq$ 12 and 
all ellipticals and lenticulars (E, S0, S0a) with $K_S$$_{tot}$ $\leq$ 8.7.\\

3) {\bf High galactic latitude}: To minimize galactic cirrus contamination, galaxies have been selected at high galactic 
latitude ($b$ $>$ $+$ 55$^\circ$) and in low galactic extinction regions 
($A_B$ $<$ 0.2; Schlegel et al. 1998). \\

\begin{figure*}
\epsscale{1.5} 
\plottwo{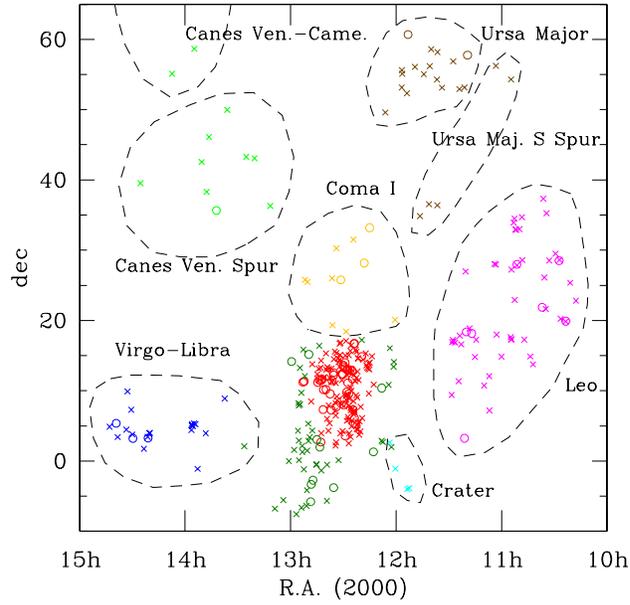}{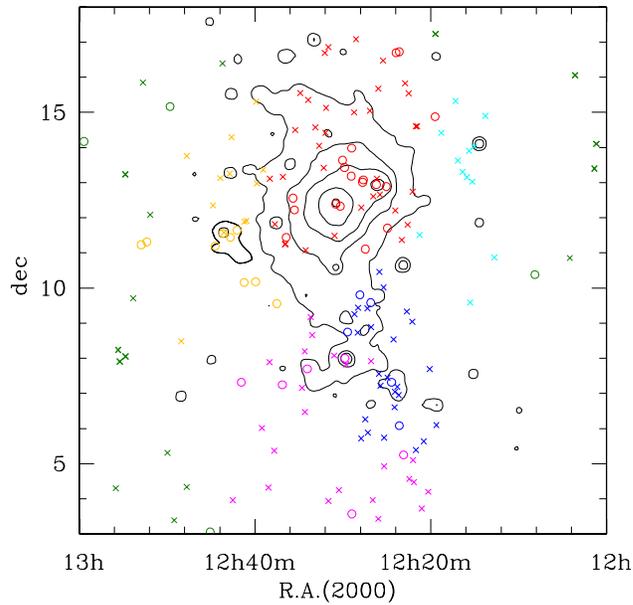}
\small{\caption{The sky distribution of the HRS for early-type (E-S0-S0a; circles)
and late-type (Sa-Sd-Im-BCD; crosses) galaxies (left panel). Dashed contours delimit the different clouds.
The big concentration of galaxies in the centre of the figure is the Virgo cluster (red colour) with its outskirts (dark green). 
Orange symbols are for Coma I Cloud, magenta for Leo Cloud, brown for 
Ursa Major Southern Spur and Cloud, cyan for Crater Cloud, light green for Canes Venatici Spur and Camelopardalis
and blue for Virgo-Lybra Cloud galaxies respectively.
The Virgo cluster region (right panel): black contours show the diffuse X-ray emission of the cluster (from B\"ohringer et al. 1994).
Red symbols are for galaxies belonging to the Virgo A cloud, blue to Virgo B, orange to Virgo E, magenta to Virgo S, cyan to Virgo N and
dark green to the Virgo outskirts, as defined by Gavazzi et al. (1999a).
\label{coord}}}
\end{figure*}

The resulting sample is composed of 323 galaxies located in the sky region 
between 10h17m $<$ R.A.(2000) $<$ 14h43m and -6$^\circ$ $<$ dec $<$ 60$^\circ$ (see Fig. \ref{coord}), 
of which 65 are early-type (E, S0 and S0a) and 258 are late-type (Sa-Sd-Im-BCD) (see Fig. \ref{stat}).
Figure \ref{stat} shows that the only galaxies which are clearly undersampled
are blue compact galaxies and dwarf irregulars, the most numerous galaxies in
the nearby universe. They will be the targets of another SPIRE key program 
\footnote{The ISM in Low Metallicity Environments: Bridging the Gap
Between Local Universe and Primordial Galaxies}.
As selected, the sample spans a large range in environment since it includes the Virgo cluster, many
galaxy groups and pairs as well as relatively isolated objects (Fig. \ref{coord}). Using the Virgo cluster 
membership criteria defined in Gavazzi et al. (1999a), the HRS includes 82 members of cluster A and B (Fig. \ref{stat})
The other galaxies are members of nearby clouds such as Leo, Ursa Major and Ursa Major Southern Spur, Crater, Coma I, Canes Venatici Spur
and Canes Venatici - Camelopardalis and Virgo-Libra Clouds (Tully 1988). As defined, the sample is thus ideal for environmental studies
(Cortese \& Hughes 2009; Hughes \& Cortese 2009).
The whole sample is given in Table \ref{TabHRS} with columns arranged as follow:\\

\noindent
Column 1: Herschel Reference Sample (HRS) name. This is the position of the galaxy in the sample list when sorted by increasing right ascension.\\
Column 2: Zwicky name, from the Catalogue of Galaxies and of Cluster of Galaxies (CGCG; Zwicky et al. 1961-1968).\\
Column 3: Virgo Cluster Catalogue (VCC) name, from Binggeli et al. (1985).\\
Column 4: Uppsala General Catalog (UGC) name (Nilson 1973).\\
Column 5: New General Catalogue (NGC) name.\\
Column 6: Index Catalogue (IC) name.\\
Columns 7 and 8: J2000 right ascension and declinations, taken from NED.\\
Column 9: Morphological type, taken from NED, or from our own classification if not available.\\
Column 10: Total K band magnitude ($K_S$$_{tot}$), from 2MASS (Skrutskie et al. 2006).\\
Column 11: Optical isophotal diameter (25 mag arcsec$^{-2}$), from NED.\\
Column 12: Heliocentric radial velocity (in km s$^{-1}$), from HI data when available, otherwise from NED.\\
Column 13: Cluster or cloud membership, from Gavazzi et al. (1999a) for Virgo and Tully (1988) or Nolthenius (1993) whenever available, 
or from our own estimate otherwise. \\
Column 14: Pair/group membership, from Karachentsev et al. (1972) or from NED whenever available, or
from our own estimate elsewhere. Pair/group membership has been assigned according to the following criteria: 
close pairs (CP) are those with a nearby companion at a distance
less than 50 kpc and a difference in redshift $<$ 600 km s$^{-1}$, as in Gavazzi et al. (1999b), while
pairs (P) up to 150 kpc. The number indicates whether the galaxy belongs to a triplet (3), quadruplet (4) and quintuplet (5). 
Groups outside Virgo and its immediate surroundings
have been determined by counting the number of bright galaxies\footnote{For bright galaxies
we intend those included in major catalogs such as NGC, UGC, IC or CGCG.} within 25 arcmin 
(which, at a median distance of 20 Mpc, corresponds to $\sim$ 150 kpc) and 600 km s$^{-1}$. Pairs in the Virgo region
are only those cataloged by Karachentsev et al. (1972).\\
Column 15: Galactic extinction $A_B$ (Schlegel et al. 1998).\\
Column 16: Alternative names.\\

\clearpage

\begin{table*}
\tiny
\caption{The Herschel Reference Sample.}
\label{TabHRS}
\[
\begin{tabular}{p{0.2cm}r p{0.2cm}p{0.2cm}p{0.2cm}p{0.2cm}cc p{2.6cm}rrrc p{0.2cm}p{0.2cm}c}
\hline
\noalign{\smallskip}
     HRS&    CGCG &      VCC &    UGC &      NGC&    IC   & RA(2000) &   dec         & type  &$K_S$ $_{tot}$& $D(25)$ &    vel     &   memb  & group & AB  & Other name\\
	&	  &	     &	      &         &         & h m s    & $^{\circ}$ ' "&       &mag           & '       &  km s$^{-1}$&         &       & mag &   \\
\noalign{\smallskip}    
\hline
\noalign{\smallskip}     
       1&   123035&        - &       -&        -&        -&101739.66& 224835.9&Pec  			&11.59& 1.00&   1175&Leo Cl.		      & 	 &  0.13& \\ 
       2&   124004&        - &    5588&        -&        -&102057.13& 252153.4&S?  			&11.03& 0.52&   1291&Leo Cl.		      & 	 &  0.10& \\ 
       3&    94026&        - &    5617&     3226&        -&102327.01& 195354.7&E2:pec;LINER;Sy3		& 8.57& 3.16&   1169&Leo Cl.		      &  CP	 &  0.10&KPG234A, ARP94 \\ 
       4&    94028&        - &    5620&     3227&        -&102330.58& 195154.2&SAB(s)pec;Sy1.5   	& 7.64& 5.37&   1148&Leo Cl.		      &  CP	 &  0.10&KPG234B, ARP94 \\ 
       5&    94052&        - &       -&        -&      610&102628.37& 201341.5&Sc   			& 9.94& 1.86&   1170&Leo Cl.		      & 	 &  0.09& \\ 
       6&   154016&        - &    5662&    3245A&        -&102701.16& 283821.9&SB(s)b  			&11.83& 3.31&   1322&Leo Cl.		      & CP	 &  0.12&NGC 3245A \\ 
       7&   154017&        - &    5663&     3245&        -&102718.39& 283026.6&SA(r)0:?;HII;LINER	& 7.86& 3.24&   1314&Leo Cl.		      & CP	 &  0.11& \\ 
       8&   154020&        - &    5685&     3254&        -&102919.92& 292929.2&SA(s)bc;Sy2   		& 8.80& 5.01&   1356&Leo Cl.		      & 	 &  0.09& \\ 
       9&   154026&        - &    5731&     3277&        -&103255.45& 283042.2&SA(r)ab;HII   		& 8.93& 1.95&   1415&Leo Cl.		      & 	 &  0.11& \\ 
      10&   183028&        - &    5738&        -&        -&103429.82& 351524.4&S?		   	&11.31& 0.91&   1516&Leo Cl.		      & 	 &  0.12& \\ 
      11&   124038&        - &    5742&     3287&        -&103447.31& 213854.0&SB(s)d   		& 9.78& 2.09&   1325&Leo Cl.		      & 	 &  0.10& \\ 
      12&   124041&        - &       -&        -&        -&103542.07& 260733.7&cI   			&11.98& 0.59&   1392&Leo Cl.		      & 	 &  0.10& \\ 
      13&   183030&        - &    5753&     3294&        -&103616.25& 371928.9&SA(s)c   		& 8.38& 3.55&   1573&Leo Cl.		      & 	 &  0.08& \\ 
      14&   124045&        - &    5767&     3301&        -&103656.04& 215255.7&(R')SB(rs)0/a   		& 8.52& 3.55&   1341&Leo Cl.		      & 	 &  0.10& \\ 
      15&    65087&        - &    5826&     3338&        -&104207.54& 134449.2&SA(s)c   		& 8.13& 5.89&   1300&Leo Cl.		      & 3	 &  0.14& \\ 
      16&    94116&        - &    5842&     3346&        -&104338.91& 145218.7&SB(rs)cd   		& 9.59& 2.69&   1260&Leo Cl.		      & 	 &  0.12& \\ 
      17&    95019&        - &    5887&     3370&        -&104704.05& 171625.3&SA(s)c   		& 9.43& 3.16&   1281&Leo Cl.		      & 	 &  0.13& \\ 
      18&   155015&        - &    5906&     3380&        -&104812.17& 283606.5&(R')SBa?   		& 9.92& 1.70&   1604&Leo Cl.		      & 	 &  0.11& \\ 
      19&   184016&        - &    5909&     3381&        -&104824.82& 344241.1&SB pec   		&10.32& 2.04&   1630&Leo Cl.		      & 	 &  0.09& \\ 
      20&   184018&        - &    5931&     3395&     2613&104950.11& 325858.3&SAB(rs)cd pec:   	& 9.95& 2.09&   1617&Leo Cl.		      &  CP	 &  0.11&KPG249A, ARP270 \\ 
      21&   155028&        - &    5958&        -&        -&105115.81& 275054.9&Sbc   			&11.56& 1.45&   1182&Leo Cl.		      &  3	 &  0.11& \\ 
      22&   155029&        - &    5959&     3414&        -&105116.23& 275830.0&S0 pec;LINER   		& 7.98& 3.55&   1414&Leo Cl.		      &  3	 &  0.11& \\ 
      23&   184028&        - &    5972&     3424&        -&105146.33& 325402.7&SB(s)b:?;HII   		& 9.04& 2.82&   1501&Leo Cl.		      &  4	 &  0.10& \\ 
      24&   184029&        - &    5982&     3430&        -&105211.41& 325701.5&SAB(rs)c   		& 8.90& 3.98&   1585&Leo Cl.		      &  4	 &  0.10& \\ 
      25&   125013&        - &    5995&     3437&        -&105235.75& 225602.9&SAB(rs)c:  		& 8.88& 2.51&   1277&Leo Cl.		      & 	 &  0.08& \\ 
      26&   184031&        - &    5990&        -&        -&105238.34& 342859.3&Sab   			&11.71& 1.35&   1569&Leo Cl.		      & 	 &  0.08& \\ 
      27&   184034&        - &    6001&     3442&        -&105308.11& 335437.3&Sa?   			&10.90& 0.62&   1734&Leo Cl.		      & 	 &  0.08& \\ 
      28&   155035&        - &    6023&     3451&        -&105420.86& 271422.9&Sd   			&10.23& 1.70&   1332&Leo Cl.		      & 	 &  0.09& \\ 
      29&    95060&        - &    6026&     3454&        -&105429.45& 172038.3&SB(s)c? sp;HII   	&10.67& 2.09&   1101&Leo Cl.		      &  4	 &  0.15&KPG257A \\ 
      30&    95062&        - &    6028&     3455&        -&105431.07& 171704.7&(R')SAB(rs)b   		&10.39& 2.38&   1105&Leo Cl.		      &  4	 &  0.14&KPG257B \\ 
      31&   267027&        - &    6024&     3448&        -&105439.24& 541818.8&I0   			& 9.47& 5.62&   1374&Ursa Maj. S S	      &  CP	 &  0.05&ARP205 \\ 
      32&    95065&        - &    6030&     3457&        -&105448.63& 173716.3&S?   			& 9.64& 0.91&   1158&Leo Cl.		      &  4	 &  0.13& \\ 
      33&    95085&        - &    6077&     3485&        -&110002.38& 145029.7&SB(r)b:   		& 9.46& 2.10&   1432&Leo Cl.		      & 	 &  0.09& \\ 
      34&    95097&        - &    6116&     3501&        -&110247.32& 175922.2&Scd   			& 9.41& 3.89&   1130&Leo Cl.		      &  P	 &  0.10&KPG263A \\ 
      35&   267037&        - &    6115&     3499&        -&110311.03& 561318.2&I0   			&10.23& 0.81&   1522&Ursa Maj. S S	      &		 &  0.04& \\ 
      36&   155049&        - &    6118&     3504&        -&110311.21& 275821.0&(R)SAB(s)ab;HII   	& 8.27& 2.69&   1536&Leo Cl.		      & P	 &  0.12& \\ 
      37&   155051&        - &    6128&     3512&        -&110402.98& 280212.5&SAB(rs)c   		& 9.65& 1.62&   1373&Leo Cl.		      & P	 &  0.12& \\ 
      38&    38129&        - &    6167&     3526&        -&110656.63& 071026.1&SAc pec sp 		&10.69& 1.91&   1419&Leo Cl.		      & 	 &  0.14& \\ 
      39&    66115&        - &    6169&        -&        -&110703.35& 120336.2&Sb:   			&11.13& 1.86&   1557&Leo Cl.		      & 	 &  0.07& \\ 
      40&    67019&        - &    6209&     3547&        -&110955.94& 104315.0&Sb:   			&10.44& 1.91&   1584&Leo Cl.		      & P	 &  0.10& \\ 
      41&    96011&        - &    6267&     3592&        -&111427.25& 171536.5&Sc? sp   		&10.78& 1.78&   1303&Leo Cl.		      & 	 &  0.07& \\ 
      42&    96013&        - &    6277&     3596&        -&111506.21& 144713.5&SAB(rs)c   		& 8.70& 4.06&   1193&Leo Cl.		      & 	 &  0.10& \\ 
      43&    96022&        - &    6299&     3608&        -&111658.96& 180854.9&E2;LINER:   		& 8.10& 3.16&   1108&Leo Cl.		      & 5	 &  0.09&KPG278B \\ 
      44&    96026&        - &    6320&        -&        -&111817.24& 185049.0&S?   			&10.99& 0.89&   1121&Leo Cl.		      & CP	 &  0.10& \\ 
      45&   291054&        - &    6330&     3619&        -&111921.60& 574527.8&(R)SA(s)0+:   		& 8.58& 2.69&   1544&Ursa Major Cl.	      & 5	 &  0.08& \\ 
      46&    96029&        - &    6343&     3626&        -&112003.80& 182124.5&(R)SA(rs)0+   		& 8.16& 2.69&   1494&Leo Cl.		      & CP	 &  0.09& \\ 
      47&   156064&        - &    6352&     3629&        -&112031.82& 265748.2&SA(s)cd:   		&10.50& 2.29&   1507&Leo Cl.		      & 	 &  0.08& \\ 
      48&   268021&        - &    6360&     3631&        -&112102.85& 531011.0&SA(s)c   		& 7.99& 5.01&   1155&Ursa Major Cl.	      & 	 &  0.07& \\ 
      49&    39130&        - &    6368&     3640&        -&112106.85& 031405.4&E3   			& 7.52& 3.98&   1251&Leo Cl.		      & 4	 &  0.19& \\ 
      50&    96037&        - &    6396&     3655&        -&112254.62& 163524.5&SA(s)c:;HII   		& 8.83& 1.55&   1500&Leo Cl.		      & 	 &  0.11& \\ 
      51&    96038&        - &    6405&     3659&        -&112345.49& 174906.8&SB(s)m?   		&10.28& 2.09&   1299&Leo Cl.		      & 	 &  0.08& \\ 
      52&   268030&        - &    6406&     3657&        -&112355.57& 525515.5&SAB(rs)c pec   		&10.29& 1.45&   1204&Ursa Major Cl.	      & 	 &  0.07& \\ 
      53&    67071&        - &    6420&     3666&        -&112426.07& 112032.0&SA(rs)c:   		& 9.23& 4.37&   1060&Leo Cl.		      & 	 &  0.14& \\ 
      54&    96045&        - &    6445&     3681&        -&112629.80& 165147.5&SAB(r)bc;LINER   	& 9.79& 2.25&   1244&Leo Cl.		      & 4	 &  0.11& \\ 
      55&    96047&        - &    6453&     3684&        -&112711.18& 170149.0&SA(rs)bc;HII   		& 9.28& 2.89&   1158&Leo Cl.		      & 4	 &  0.11& \\ 
      56&   291072&        - &    6458&     3683&        -&112731.85& 565237.4&SB(s)c?;HII   		& 8.67& 1.86&   1708&Ursa Major Cl.	      & 3	 &  0.07& \\ 
      57&    96049&        - &    6460&     3686&        -&112743.95& 171326.8&SB(s)bc   		& 8.49& 3.19&   1156&Leo Cl.		      & 4	 &  0.10& \\ 
      58&    96050&        - &    6464&     3691&        -&112809.41& 165513.7&SBb?   			&10.51& 1.35&   1067&Leo Cl.		      & 4	 &  0.11& \\ 
      59&    67084&        - &    6474&     3692&        -&112824.01& 092427.5&Sb;LINER;HII   		& 9.52& 3.16&   1717&Leo Cl.		      & 	 &  0.14& \\ 
      60&   268051&        - &    6547&     3729&        -&113349.34& 530731.8&SB(r)a pec   		& 8.73& 2.82&	991&Ursa Major Cl.	      & P	 &  0.05&KPG209B \\ 
      61&   292009&        - &    6575&        -&        -&113626.47& 581129.0&Scd:;HII   		&11.40& 1.95&   1217&Ursa Major Cl.	      & 4	 &  0.07& \\ 
      62&   186012&        - &    6577&     3755&        -&113633.37& 362437.2&SAB(rs)c pec   		&10.60& 3.16&   1571&Ursa Maj. S S	      &	 	 &  0.10& \\ 
      63&   268063&        - &    6579&     3756&        -&113648.02& 541736.8&SAB(rs)bc   		& 8.78& 4.17&   1289&Ursa Major Cl.	      & 	 &  0.05& \\ 
      64&   292017&        - &    6629&     3795&        -&114006.84& 583647.2&Sc;HII   		&10.64& 2.14&   1213&Ursa Major Cl.	      & 3	 &  0.06& \\ 
      65&   292019&        - &    6640&     3794&        -&114053.42& 561207.3&SAB(s)d 			&11.60& 2.24&   1383&Ursa Major Cl.	      & 	 &  0.06& \\ 
      66&   186024&        - &    6651&     3813&        -&114118.65& 363248.3&SA(rs)b:   		& 8.86& 2.24&   1468&Ursa Maj. S S	      &		 &  0.08& \\ 
      67&   268076&        - &    6706&        -&        -&114414.83& 550205.9&SB(s)m: 			&11.28& 1.91&   1436&Ursa Major Cl.	      & 	 &  0.06&NGC 3846A \\ 
      68&   186045&        - &       -&        -&        -&114625.96& 345109.2&S?   			&11.44& 0.32&   1412&Ursa Maj. S S	      &		 &  0.09&MRK 429 \\ 
      69&   268088&        - &    6787&     3898&        -&114915.37& 560503.7&SA(s)ab;LINE;HII   	& 7.66& 4.37&   1171&Ursa Major Cl.	      & 	 &  0.09& \\ 
      70&        -&        - &       -&        -&     2969&115231.27&-035220.1&SB(r)bc?;HII 		&11.15& 1.23&   1617&Crater Cl.  	      & 3	 &  0.12& \\ 
      71&   292042&        - &    6860&     3945&        -&115313.73& 604032.0&SB(rs)0+;LINER   	& 7.53& 5.25&   1259&Ursa Major Cl.	      & 	 &  0.12& \\ 
      72&        -&        - &       -&     3952&     2972&115340.63&-035947.5&IBm: sp;HII 		&11.01& 1.58&   1577&Crater Cl.  	      & 3	 &  0.11& \\ 
      73&   269013&        - &    6870&     3953&        -&115348.92& 521936.4&SB(r)bc;HII/LINER   	& 7.05& 6.92&   1050&Ursa Major Cl.	      & P	 &  0.13& \\ 
      74&   269019&        - &    6918&     3982&        -&115628.10& 550730.6&SAB(r)b:;HII;Sy2   	& 8.85& 2.34&   1108&Ursa Major Cl.	      & 4	 &  0.06& \\ 
      75&   269020&        - &    6919&        -&        -&115637.51& 553759.5&Sdm:   			&11.56& 1.45&   1283&Ursa Major Cl.	      & 4	 &  0.06& \\ 
      76&   269022&        - &    6923&        -&        -&115649.43& 530937.3&Im:   			&11.32& 2.00&   1069&Ursa Major Cl.           & 4	 &  0.12& \\ 
      77&    13033&        - &    6993&     4030&        -&120023.64&-010600.0&SA(s)bc;HII   		& 7.33& 4.17&   1458&Crater Cl.  	      & P	 &  0.11& \\ 
      78&    98019&        - &    6995&     4032&        -&120032.82& 200426.0&Im:   			&10.45& 1.86&   1269&Coma I Cl.  	      & 	 &  0.15& \\ 
      79&    69024&        - &    7001&     4019&      755&120110.39& 140616.2&SBb? sp   		&11.33& 2.40&   1508&Virgo Out.  	      &		 &  0.14& \\ 
      80&    69027&        - &    7002&     4037&        -&120123.67& 132403.7&SB(rs)b:   		&10.11& 2.51&	932&Virgo Out.  	      &		 &  0.12& \\ 
      81&    13046&        - &    7021&     4045&        -&120242.26& 015836.4&SAB(r)a;HII   		& 8.75& 3.00&   2011&Virgo Out.  	      &		 &  0.10& \\ 
      82&    98037&        - &       -&        -&        -&120335.94& 160320.0&Sab   			&11.19& 0.60&	931&Virgo Out.  	      &		 &  0.13&KUG1201+163 \\ 
      83&    41031&        - &    7035&        -&        -&120340.14& 023828.4&SB(r)a:;HII   		&11.82& 1.10&   1232&Crater Cl.  	      & 	 &  0.12& \\ 
      84&    69036&        - &    7048&     4067&        -&120411.55& 105115.8&SA(s)b:   		& 9.90& 1.20&   2424&Virgo Out.  	      &		 &  0.11& \\ 
      85&   243044&        - &    7095&     4100&        -&120608.60& 493456.3&(R')SA(rs)bc;HII    	& 8.03& 5.37&   1072&Ursa Major Cl.	      & 	 &  0.10& \\ 
      86&    41041&        - &    7111&     4116&        -&120736.82& 024132.0&SB(rs)dm   		&10.27& 3.80&   1309&Virgo Out.  	      &  P	 &  0.10&KPG322A \\ 
      87&    69058&        - &    7117&     4124&        -&120809.64& 102243.4&SA(r)0+   		& 8.49& 4.10&   1652&Virgo Out.  	      &		 &  0.12& \\ 
      88&    41042&        - &    7116&     4123&        -&120811.11& 025241.8&SB(r)c;Sbrst;HII   	& 8.79& 5.00&   1326&Virgo Out.  	      &  P	 &  0.09&KPG322B \\ 
      89&    69088&       66 &    7215&     4178&        -&121246.45& 105157.5&SB(rs)dm;HII   		& 9.58& 5.35&	369&Virgo N Cl. 	      & 	 &  0.12& \\ 
      90&    13104&        - &    7214&     4179&        -&121252.11& 011758.9&Sb(f)   			& 7.92& 3.80&   1279&Virgo Out.  	      &		 &  0.14& \\ 
\noalign{\smallskip}
\hline
\end{tabular}
\]
\end{table*}

\addtocounter{table}{-1}
\begin{table*}
\tiny
\caption{continue}
\label{Tabmod}
\[
\begin{tabular}{p{0.2cm}r p{0.2cm}p{0.2cm}p{0.2cm}p{0.2cm}cc p{2.6cm}rrrc p{0.2cm}p{0.2cm}c}
\hline
\noalign{\smallskip}
     HRS&    CGCG &      VCC &    UGC &      NGC&    IC   & RA(2000) &   dec         & type  &$K_S$ $_{tot}$& $D(25)$ &    vel     &   memb  & group & AB  & Other name\\
	&	  &	     &	      &         &         & h m s    & $^{\circ}$ ' "&       &mag           & '       &  km s$^{-1}$&         &       & mag &   \\
\noalign{\smallskip}    
\hline
\noalign{\smallskip}     
      91&    98108&       92 &    7231&     4192&        -&121348.29& 145401.2&SAB(s)ab;HII;Sy   	& 6.89& 9.78&   -135&Virgo N Cl. 	  &		&  0.15  &M 98 \\
      92&    69101&      131 &    7255&        -&     3061&121504.44& 140144.3&SBc? sp 			&10.64& 2.60&   2317&Virgo N Cl. 	  &		&  0.16  & \\
      93&   187029&        - &    7256&     4203&        -&121505.06& 331150.4&SAB0-:;LINER;Sy3   	& 7.41& 3.39&   1091&Coma I Cl.  	  &		&  0.05  & \\
      94&    69104&      145 &    7260&     4206&        -&121516.81& 130126.3&SA(s)bc:   		& 9.39& 5.10&	702&Virgo N Cl. 	  &		&  0.14  & \\
      95&    69107&      152 &    7268&     4207&        -&121530.50& 093505.6&Scd   			& 9.44& 1.96&	592&Virgo N Cl. 	  &		&  0.07  & \\
      96&    69110&      157 &    7275&     4212&        -&121539.36& 135405.4&SAc:;HII 		& 8.38& 3.60&	-83&Virgo N Cl. 	  &		&  0.14  & \\
      97&    69112&      167 &    7284&     4216&        -&121554.44& 130857.8&SAB(s)b:;HII/LINER   	& 6.52& 9.12&	140&Virgo N Cl. 	  &		&  0.14  & \\
      98&    69119&      187 &    7291&     4222&        -&121622.52& 131825.5&Sc   			&10.33& 3.52&	226&Virgo N Cl. 	  &		&  0.14  & \\
      99&    69123&      213 &    7305&        -&     3094&121656.00& 133731.0&S;BCD   			&11.25& 0.93&   -162&Virgo N Cl. 	  &		&  0.15  & \\
     100&    98130&      226 &    7315&     4237&        -&121711.42& 151926.3&SAB(rs)bc;HII   		&10.03& 2.01&	864&Virgo N Cl. 	  &		&  0.13  & \\
     101&   158060&        - &    7338&     4251&        -&121808.31& 281031.1&SB0? sp   		& 7.73& 3.63&   1014&Coma I Cl.  	  &		&  0.10  & \\
     102&    98144&      307 &    7345&     4254&        -&121849.63& 142459.4&SA(s)c   		& 6.93& 6.15&   2405&Virgo N Cl. 	  &		&  0.17  &M 99 \\
     103&    42015&      341 &    7361&     4260&        -&121922.24& 060555.2&SB(s)a   		& 8.54& 3.52&   1935&Virgo B		  &		&  0.10  & \\
     104&    99015&        - &    7366&        -&        -&121928.66& 171349.4&Spiral   		&11.99& 1.20&	925&Virgo Out.  	  &		&  0.11  & \\
     105&    99014&      355 &    7365&     4262&        -&121930.58& 145239.8&SB(s)0-?   		& 8.36& 1.87&   1369&Virgo A		  &		&  0.15  & \\
     106&    42032&      393 &    7385&     4276&        -&122007.50& 074131.2&S(s)c II   		&10.69& 2.10&   2617&Virgo B		  &		&  0.12  & \\
     107&    42033&      404 &    7387&        -&        -&122017.35& 041205.1&Sd(f) 			&10.74& 1.89&   1733&Virgo S Cl. 	  &		&  0.10  & \\
     108&    42037&      434 &       -&     4287&        -&122048.49& 053823.5&Sc(f)   			&11.02& 1.76&   2155&Virgo B		  &		&  0.08  & \\
     109&    42038&      449 &    7403&     4289&        -&122102.25& 034319.7&SA(s)cd: sp 		& 9.89& 4.33&   2541&Virgo S Cl. 	  &		&  0.09  & \\
     110&    70024&      465 &    7407&     4294&        -&122117.79& 113040.0&SB(s)cd   		& 9.70& 3.95&	357&Virgo N Cl. 	  &CP		&  0.15  &KPG330B \\
     111&    99024&      483 &    7412&     4298&        -&122132.76& 143622.2&SA(rs)c   		& 8.47& 3.60&   1136&Virgo A		  &CP		&  0.15  &KPG332A \\
     112&    42044&      492 &    7413&     4300&        -&122141.47& 052305.4&Sa   			& 9.53& 2.16&   2310&Virgo B		  &		&  0.09  & \\
     113&    99027&      497 &    7418&     4302&        -&122142.48& 143553.9&Sc: sp   		& 7.83& 6.74&   1150&Virgo A		  &CP		&  0.15  &KPG332B \\
     114&    42045&      508 &    7420&     4303&        -&122154.90& 042825.1&SAB(rs)bc;HII;Sy2   	& 6.84& 6.59&   1568&Virgo S Cl. 	  &		&  0.10  &M 61 \\
     115&    42047&      517 &    7422&        -&        -&122201.30& 050600.2&SBab(s)   		&10.79& 1.41&   1864&Virgo S Cl. 	  &		&  0.08  & \\
     116&    70031&      522 &    7432&     4305&        -&122203.60& 124427.3&SA(r)a   		& 9.83& 2.60&   1888&Virgo A		  &CP		&  0.18  &KPG333A \\
     117&    70029&      524 &    7431&     4307&        -&122205.63& 090236.8&Sb   			& 8.72& 3.95&   1035&Virgo B		  &		&  0.10  & \\
     118&    42053&      552 &    7439&        -&        -&122227.25& 043358.7&SAB(s)cd 		&11.20& 1.89&   1296&Virgo S Cl. 	  &		&  0.10  & \\
     119&    99029&      559 &    7442&     4312&        -&122231.36& 153216.5&SA(rs)ab: sp 		& 8.79& 5.10&	153&Virgo A		  &		&  0.12  & \\
     120&    70034&      570 &    7445&     4313&        -&122238.55& 114803.4&SA(rs)ab: sp   		& 8.47& 5.10&   1443&Virgo A		  &		&  0.16  & \\
     121&    70035&      576 &    7447&     4316&        -&122242.24& 091956.9&Sbc     			& 9.25& 2.48&   1254&Virgo B		  &		&  0.10  & \\
     122&    99030&      596 &    7450&     4321&        -&122254.90& 154920.6&SAB(s)bc;LINER;HII   	& 6.59& 9.12&   1575&Virgo A		  &		&  0.11  &M 100 \\
     123&    42063&      613 &    7451&     4324&        -&122306.18& 051501.5&SA(r)0+   		& 8.48& 3.52&   1670&Virgo S Cl. 	  &		&  0.10  & \\
     124&    70039&      630 &    7456&     4330&        -&122317.25& 112204.7&Scd   			& 9.51& 5.86&   1564&Virgo A		  &		&  0.11  & \\
     125&    42068&      648 &    7461&     4339&        -&122334.94& 060454.2&E0;Sy2   		& 8.54& 2.31&   1298&Virgo B		  &		&  0.11  & \\
     126&    99036&      654 &    7467&     4340&        -&122335.31& 164319.9&SB(r)0+   		& 8.32& 3.60&	930&Virgo A		  &		&  0.11  & \\
     127&    42070&      656 &    7465&     4343&        -&122338.70& 065714.7&SA(rs)b:  		& 8.97& 2.48&   1014&Virgo B		  &		&  0.09  & \\
     128&    42072&      667 &    7469&        -&     3259&122348.52& 071112.6&SAB(s)dm?   		&11.06& 1.89&   1420&Virgo B		  &		&  0.10  & \\
     129&    99038&      685 &    7473&     4350&        -&122357.81& 164136.1&SA0;LINER   		& 7.82& 3.20&   1241&Virgo A		  &		&  0.12  & \\
     130&    70045&      692 &    7476&     4351&        -&122401.56& 121218.1&SB(rs)ab pec:   		&10.24& 2.92&   2324&Virgo A		  &		&  0.13  & \\
     131&    42079&      697 &    7474&        -&     3267&122405.53& 070228.6&SA(s)cd 			&10.95& 1.55&   1231&Virgo B		  &		&  0.10  & \\
     132&    42080&      699 &    7477&        -&     3268&122407.44& 063626.9&Sm/Im   			&11.49& 1.95&	727&Virgo B		  &		&  0.11  & \\
     133&   158099&        - &    7483&     4359&        -&122411.06& 313117.8&SB(rs)c? sp   		&10.81& 3.60&   1253&Coma I Cl.  	  &		&  0.11  & \\
     134&    70048&      713 &    7482&     4356&        -&122414.53& 083208.9&Sc   			& 9.69& 3.20&   1137&Virgo B		  &		&  0.12  & \\
     135&    42083&      731 &    7488&     4365&        -&122428.23& 071903.1&E3   			& 6.64& 8.73&   1240&Virgo B		  &		&  0.09  & \\
     136&    42089&      758 &    7492&     4370&        -&122454.93& 072640.4&Sa   			& 9.31& 1.76&	784&Virgo B		  &		&  0.10  & \\
     137&    70057&      759 &    7493&     4371&        -&122455.43& 114215.4&SB(r)0+   		& 7.72& 5.10&	943&Virgo A		  &		&  0.16  & \\
     138&    70058&      763 &    7494&     4374&        -&122503.78& 125313.1&E1;LERG;LINER;Sy2   	& 6.22&10.07&	910&Virgo A		  &		&  0.17  &M 84 \\
     139&    42093&      787 &    7498&     4376&        -&122518.06& 054428.3&Im   			&11.23& 1.84&   1136&Virgo B		  &		&  0.10  & \\
     140&    42092&      785 &    7497&     4378&        -&122518.09& 045530.2&(R)SA(s)a;Sy2   		& 8.51& 3.06&   2557&Virgo S Cl. 	  &		&  0.07  & \\
     141&    70061&      792 &    7503&     4380&        -&122522.17& 100100.5&SA(rs)b:?   		& 8.33& 3.52&	971&Virgo B		  &		&  0.10  & \\
     142&    99044&      801 &    7507&     4383&        -&122525.50& 162812.0&Sa? pec;HII   		& 9.49& 2.60&   1710&Virgo A		  &		&  0.10  & \\
     143&    42095&      827 &    7513&        -&        -&122542.63& 071300.1&SB(s)cd: sp   		& 9.79& 3.60&	992&Virgo B		  &		&  0.11  &IC 3322A\\
     144&    70068&      836 &    7520&     4388&        -&122546.82& 123943.5&SA(s)b: sp;Sy2   	& 8.00& 5.10&   2515&Virgo A		  &		&  0.14  & \\
     145&    70067&      849 &    7519&     4390&        -&122550.67& 102732.6&Sbc(s) II 		&10.33& 2.18&   1103&Virgo B		  &		&  0.13  & \\
     146&    42098&      851 &    7518&        -&     3322&122554.12& 073317.4&SAB(s)cd: sp   		&10.47& 2.16&   1195&Virgo B		  &		&  0.12  & \\
     147&    42099&      859 &    7522&        -&        -&122558.30& 032547.3&Sd(f)   			&10.18& 2.92&   1428&Virgo S Cl. 	  &		&  0.10  & \\
     148&    99049&      865 &    7526&     4396&        -&122558.80& 154017.3&SAd: sp 			&10.34& 3.36&   -124&Virgo A		  &		&  0.11  & \\
     149&    70071&      873 &    7528&     4402&        -&122607.56& 130646.0&Sb   			& 8.49& 3.95&	234&Virgo A		  &		&  0.12  & \\
     150&    70072&      881 &    7532&     4406&        -&122611.74& 125646.4&S0(3)/E3   		& 6.10&11.37&   -221&Virgo A		  &		&  0.13  &M 86 \\
     151&    70076&      912 &    7538&     4413&        -&122632.25& 123639.5&(R')SB(rs)ab:   		& 9.80& 2.92&	105&Virgo A		  &		&  0.14  & \\
     152&    42104&      921 &    7536&     4412&        -&122636.10& 035752.7&SB(r)b? pec;LINER   	& 9.65& 1.89&   2289&Virgo S Cl. 	  &		&  0.08  & \\
     153&    42105&      938 &    7541&     4416&        -&122646.72& 075508.4&SB(rs)cd:;Sbrst   	&10.97& 2.18&   1395&Virgo S Cl. 	  &		&  0.11  & \\
     154&    70082&      939 &    7546&        -&        -&122647.23& 085304.6&SAB(s)cd   		&10.71& 3.45&   1271&Virgo B		  &		&  0.13  &NGC 4411B\\
     155&    70080&      944 &    7542&     4417&        -&122650.62& 093503.0&SB0: s   		& 8.17& 3.60&	832&Virgo B		  &		&  0.10  & \\
     156&    99054&      958 &    7551&     4419&        -&122656.43& 150250.7&SB(s)a;LINER;HII   	& 7.74& 3.52&   -273&Virgo A		  &		&  0.14  & \\
     157&    42106&      957 &    7549&     4420&        -&122658.48& 022939.7&SB(r)bc:   		& 9.66& 2.01&   1695&Virgo S Cl. 	  &		&  0.08  & \\
     158&    42107&      971 &    7556&     4423&        -&122708.97& 055248.6&Sdm:   			&11.05& 3.06&   1120&Virgo B		  &		&  0.09  & \\
     159&    70090&      979 &    7561&     4424&        -&122711.59& 092514.0&SB(s)a:;HII  		& 9.09& 4.33&	438&Virgo B		  &		&  0.09  & \\
     160&    42111&     1002 &    7566&     4430&        -&122726.41& 061546.0&SB(rs)b:   		& 9.35& 3.02&   1450&Virgo B		  &CP		&  0.08  &KPG338A \\
     161&    70093&     1003 &    7568&     4429&        -&122726.56& 110627.1&SA(r)0+;LINER;HII   	& 6.78& 8.12&   1130&Virgo A		  &		&  0.14  & \\
     162&    70098&     1030 &    7575&     4435&        -&122740.49& 130444.2&SB(s)0;LINER;HII   	& 7.30& 2.92&	775&Virgo A		  &		&  0.13  & \\
     163&    70097&     1043 &    7574&     4438&        -&122745.59& 130031.8&SA(s)0/a pec:;LINER      & 7.27& 8.12&	 70&Virgo A		  &		&  0.12  & \\
     164&    70099&     1047 &    7581&     4440&        -&122753.57& 121735.6&SB(rs)a   		& 8.91& 2.01&	724&Virgo A		  &		&  0.12  & \\
     165&    42117&     1048 &    7579&        -&        -&122755.39& 054316.4&Sdm:   			&11.58& 1.89&   2252&Virgo B		  &		&  0.09  & \\
     166&    70100&     1062 &    7583&     4442&        -&122803.89& 094813.0&SB(s)0   		& 7.29& 5.05&	517&Virgo B		  &		&  0.10  & \\
     167&    70104&     1086 &    7587&     4445&        -&122815.94& 092610.7&Sab: sp   		& 9.83& 3.20&	328&Virgo B		  &		&  0.11  & \\
     168&    70108&     1091 &    7590&        -&        -&122818.77& 084346.1&Sbc  			&11.77& 1.76&   1119&Virgo B		  &		&  0.09  & \\
     169&    99063&        - &    7595&        -&     3391&122827.28& 182455.1&Scd:   			&10.45& 1.10&   1701&Coma I Cl.  	  &		&  0.14  & \\
     170&    99062&     1110 &    7594&     4450&        -&122829.63& 170505.8&SA(s)ab;LINER;Sy3   	& 7.05& 6.15&   1954&Virgo A		  &		&  0.12  & \\
     171&    70111&     1118 &    7600&     4451&        -&122840.55& 091532.2&Sbc:   			& 9.99& 1.96&	865&Virgo B		  &		&  0.08  & \\
     172&    99065&     1126 &    7602&        -&     3392&122843.26& 145958.2&SAb: 			& 9.26& 2.92&   1687&Virgo A		  &		&  0.16  & \\
     173&    42124&     1145 &    7609&     4457&        -&122859.01& 033414.2&(R)SAB(s)0/a;LINER   	& 7.78& 2.92&	884&Virgo S Cl. 	  &		&  0.09  & \\
     174&    70116&     1154 &    7614&     4459&        -&122900.03& 135842.9&SA(r)0+;HII;LINER   	& 7.15& 3.36&   1210&Virgo A		  &		&  0.20  & \\
     175&    70115&     1158 &    7613&     4461&        -&122903.01& 131101.5&SB(s)0+:   		& 8.01& 3.52&   1919&Virgo A		  &		&  0.10  & \\
     176&    70121&     1190 &    7622&     4469&        -&122928.03& 084459.7&SB(s)0/a? sp   		& 8.04& 4.33&	508&Virgo B		  &		&  0.09  & \\
     177&    42132&     1205 &    7627&     4470&        -&122937.78& 074927.1&Sa?;HII   		&10.12& 1.84&   2339&Virgo S Cl. 	  &		&  0.10  & \\
     178&    42134&     1226 &    7629&     4472&        -&122946.76& 080001.7&E2/S0;Sy2;LINER   	& 5.40&10.25&	868&Virgo S Cl. 	  &		&  0.10  &M 49 \\
     179&    70125&     1231 &    7631&     4473&        -&122948.87& 132545.7&E5   			& 7.16& 4.04&   2236&Virgo A		  &		&  0.12  & \\
     180&    70129&     1253 &    7638&     4477&        -&123002.17& 133811.2&SB(s)0:?;Sy2   		& 7.35& 3.60&   1353&Virgo A		  &		&  0.14  & \\
\noalign{\smallskip}
\hline
\end{tabular}
\]
\end{table*}
\addtocounter{table}{-1}

\begin{table*}
\tiny
\caption{continue}
\label{Tabmod}
\[
\begin{tabular}{p{0.2cm}r p{0.2cm}p{0.2cm}p{0.2cm}p{0.2cm}cc p{2.6cm}rrrc p{0.2cm}p{0.2cm}c}
\hline
\noalign{\smallskip}
     HRS&    CGCG &      VCC &    UGC &      NGC&    IC   & RA(2000) &   dec         & type  &$K_S$ $_{tot}$& $D(25)$ &    vel     &   memb  & group & AB  & Other name\\
	&	  &	     &	      &         &         & h m s    & $^{\circ}$ ' "&       &mag           & '       &  km s$^{-1}$&         &       & mag &   \\
\noalign{\smallskip}    
\hline
\noalign{\smallskip}     
     181&    70133&     1279 &    7645&     4478&        -&123017.42& 121942.8&E2   			& 8.36& 1.89&	 1370&Virgo A		   &	       &  0.11   & \\ 
     182&    42139&     1290 &    7647&     4480&        -&123026.78& 041447.3&SAB(s)c   		& 9.75& 2.01&	 2438&Virgo S Cl.	   &	       &  0.10   & \\ 
     183&    70139&     1316 &    7654&     4486&        -&123049.42& 122328.0&E+0-1 pec;NLRG;Sy   	& 5.81&11.00&	 1292&Virgo A		   &	       &  0.10   &M 87 \\ 
     184&    70140&     1326 &    7657&     4491&        -&123057.13& 112900.8&SB(s)a:   		& 9.88& 1.89&	  497&Virgo A		   &	       &  0.18   & \\ 
     185&    42141&     1330 &    7656&     4492&        -&123059.74& 080440.6&SA(s)a?   		& 9.08& 1.96&	 1777&Virgo S Cl.	   &	       &  0.11   & \\ 
     186&   129005&        - &    7662&     4494&        -&123124.03& 254629.9&E1-2;LINER   		& 7.00& 4.79&	 1310&Coma I Cl.	   &	       &  0.09   & \\ 
     187&    42144&     1375 &    7668&     4505&        -&123139.21& 035622.1&SB(rs)m   		& 9.56& 4.76&	 1732&Virgo S Cl.	   &	       &  0.11   &NGC 4496A, KPG343A \\ 
     188&    99075&     1379 &    7669&     4498&        -&123139.57& 165110.1&SAB(s)d 			& 9.66& 2.85&	 1505&Virgo A		   &	       &  0.13   & \\ 
     189&    99077&     1393 &    7676&        -&      797&123154.76& 150726.2&SB(s)c II.5   		&10.80& 1.69&	 2100&Virgo A		   &	       &  0.13   & \\ 
     190&    99076&     1401 &    7675&     4501&        -&123159.22& 142513.5&SA(rs)b;HII;Sy2   	& 6.27& 7.23&	 2284&Virgo A		   &	       &  0.16   &M 88 \\ 
     191&    99078&     1410 &    7677&     4502&        -&123203.35& 164115.8&Scd:   			&11.90& 1.48&	 1629&Virgo A		   &	       &  0.13   & \\ 
     192&    70152&     1419 &    7682&     4506&        -&123210.53& 132510.6&Sa pec?   		&10.26& 2.16&	  737&Virgo A		   &	       &  0.13   & \\ 
     193&    70157&     1450 &    7695&        -&     3476&123241.88& 140301.8&IB(s)m:			&10.91& 2.60&	 -173&Virgo A		   &	       &  0.15   & \\ 
     194&    14063&        - &    7694&     4517&        -&123245.59& 000654.1&SA(s)cd: sp   		& 7.33&11.00&	 1129&Virgo Out.	   &P	       &  0.10   &KPG344B \\ 
     195&    99087&     1479 &    7703&     4516&        -&123307.56& 143429.8&SB(rs)ab?   		& 9.99& 2.16&	  958&Virgo A		   &	       &  0.16   & \\ 
     196&    70167&     1508 &    7709&     4519&        -&123330.25& 083917.1&SB(rs)d   		& 9.56& 3.60&	 1212&Virgo S Cl.	   &	       &  0.09   & \\ 
     197&    70168&     1516 &    7711&     4522&        -&123339.66& 091029.5&SB(s)cd: sp   		&10.35& 4.04&	 2330&Virgo S Cl.	   &	       &  0.09   & \\ 
     198&   159016&        - &    7714&     4525&        -&123351.19& 301639.1&Scd:   			& 9.99& 3.00&	 1174&Coma I Cl.	   &	       &  0.10   & \\ 
     199&    99090&     1532 &    7716&        -&      800&123356.66& 152117.4&SB(rs)c pec? 		&10.58& 1.96&	 2335&Virgo A		   &	       &  0.16   & \\ 
     200&    42155&     1535 &    7718&     4526&        -&123403.03& 074156.9&SAB(s)0:   		& 6.47& 7.00&	  448&Virgo S Cl.	   &	       &  0.10   & \\ 
     201&    42156&     1540 &    7721&     4527&        -&123408.50& 023913.7&SAB(s)bc;HII;LINER   	& 6.93& 5.86&	 1736&Virgo S Cl.	   &	       &  0.10   & \\ 
     202&    70173&     1549 &    7728&        -&     3510&123414.79& 110417.7&S?   			&11.42& 1.10&	 1357&Virgo A		   &	       &  0.13   &  \\ 
     203&    42158&     1554 &    7726&     4532&        -&123419.33& 062803.7&IBm;HII   		& 9.48& 2.60&	 2021&Virgo S Cl.	   &	       &  0.09   & \\ 
     204&    42159&     1555 &    7727&     4535&        -&123420.31& 081151.9&SAB(s)c;HII   		& 7.38& 8.33&	 1962&Virgo S Cl.	   &	       &  0.08   &  \\ 
     205&    14068&     1562 &    7732&     4536&        -&123427.13& 021116.4&SAB(rs)bc;HII;Sbrst   	& 7.52& 7.23&	 1807&Virgo S Cl.	   &	       &  0.08   & \\ 
     206&    42162&     1575 &    7736&        -&     3521&123439.42& 070936.0&SBm pec;BCD   		&11.01& 2.00&	  597&Virgo S Cl.	   &	       &  0.10   & \\ 
     207&    99093&     1588 &    7742&     4540&        -&123450.87& 153305.2&SAB(rs)cd;LINER;Sy1   	& 9.24& 2.60&	 1288&Virgo A		   &	       &  0.14   & \\ 
     208&    99096&     1615 &    7753&     4548&        -&123526.43& 142946.8&SBb(rs);LINER;Sy   	& 7.12& 6.00&	  484&Virgo A		   &	       &  0.16   &M 91 \\ 
     209&        -&        - &       -&     4546&        -&123529.51&-034735.5&SB(s)0-:   		& 7.39& 3.31&	 1050&Virgo Out.	   &	       &  0.15   & \\ 
     210&    70182&     1619 &    7757&     4550&        -&123530.61& 121315.4&SB0: sp;LINER 		& 8.69& 3.95&	  381&Virgo A		   &	       &  0.17   & \\ 
     211&    70184&     1632 &    7760&     4552&        -&123539.88& 123321.7&E;LINER;HII;Sy2   	& 6.73& 7.23&	  322&Virgo A		   &	       &  0.18   &M 89 \\ 
     212&    99098&        - &    7768&     4561&        -&123608.14& 191921.4&SB(rs)dm   		&10.63& 1.51&	 1410&Coma I Cl.	   &	       &  0.11   &KPG346A \\ 
     213&   129010&        - &    7772&     4565&        -&123620.78& 255915.6&SA(s)b? sp;Sy3;Sy1.9   	& 6.06&14.18&	 1233&Coma I Cl.	   &3	       &  0.07   & \\ 
     214&    70186&     1664 &    7773&     4564&        -&123626.99& 112621.5&E6   			& 7.94& 4.33&	 1165&Virgo A		   &	       &  0.15   & \\ 
     215&    70189&     1673 &    7777&     4567&        -&123632.71& 111528.8&SA(rs)bc   		& 8.30& 2.92&	 2277&Virgo A		   &CP         &  0.14   &KPG347A \\ 
     216&    70188&     1676 &    7776&     4568&        -&123634.26& 111420.0&SA(rs)bc   		& 7.52& 5.10&	 2255&Virgo A		   &CP         &  0.14   &KPG347B \\ 
     217&    70192&     1690 &    7786&     4569&        -&123649.80& 130946.3&SAB(rs)ab;LINER;Sy   	& 6.58&10.73&	 -216&Virgo A		   &	       &  0.20   &M 90 \\ 
     218&    42178&     1692 &    7785&     4570&        -&123653.40& 071448.0&S0(7)/E7   		& 7.69& 3.52&	 1730&Virgo S Cl.	   &	       &  0.10   & \\ 
     219&    70195&     1720 &    7793&     4578&        -&123730.55& 093318.4&SA(r)0:   		& 8.40& 3.77&	 2284&Virgo E Cl.	   &	       &  0.09   & \\ 
     220&    70197&     1727 &    7796&     4579&        -&123743.52& 114905.5&SAB(rs)b;LINER;Sy1.9  	& 6.49& 6.29&	 1520&Virgo A		   &	       &  0.18   &M 58 \\ 
     221&    42183&     1730 &    7794&     4580&        -&123748.40& 052206.4&SAB(rs)a pec   		& 8.77& 2.16&	 1032&Virgo S Cl.	   &	       &  0.10   & \\ 
     222&    70199&     1757 &    7803&     4584&        -&123817.89& 130635.5&SAB(s)a?   		&10.46& 1.96&	 1783&Virgo A		   &	       &  0.16   & \\ 
     223&    42186&     1758 &    7802&        -&        -&123820.82& 075328.7&Sdm   			&11.76& 1.89&	 1788&Virgo S Cl.	   &	       &  0.12   & \\ 
     224&    42187&     1760 &    7804&     4586&        -&123828.44& 041908.8&SA(s)a: sp   		& 8.47& 4.33&	  792&Virgo S Cl.	   &	       &  0.16   & \\ 
     225&    70202&     1778 &    7817&        -&     3611&123904.14& 132148.7&S?   			&11.42& 1.76&	 2750&Virgo E Cl.	   &	       &  0.15   & \\ 
     226&    42191&     1780 &    7821&     4591&        -&123912.44& 060044.3&Sb   			&10.24& 1.96&	 2424&Virgo S Cl.	   &	       &  0.09   & \\ 
     227&    14091&        - &    7819&     4592&        -&123918.74& 0-3155.2&SA(s)dm:   		&10.22& 5.75&	 1069&Virgo Out.	   &	       &  0.10   & \\ 
     228&        -&        - &       -&        -&        -&123922.26&-053953.3&Pec   			&11.95& 0.43&	 1199&Virgo Out.	   &CP         &  0.11   &LCRSB123647.4-052325 \\ 
     229&    70204&     1809 &    7825&        -&     3631&123948.02& 125826.1&S?   			&11.11& 1.10&	 2839&Virgo E Cl.	   &	       &  0.17   & \\ 
     230&    99106&     1811 &    7826&     4595&        -&123951.91& 151752.1&SAB(rs)b?   		&10.03& 2.16&	  632&Virgo E Cl.	   &	       &  0.16   & \\ 
     231&    70206&     1813 &    7828&     4596&        -&123955.94& 101033.9&SB(r)0+;LINER:   	& 7.46& 4.76&	 1834&Virgo E Cl.	   &	       &  0.10   & \\ 
     232&    70213&     1859 &    7839&     4606&        -&124057.56& 115443.6&SB(s)a:   		& 9.17& 5.10&	 1645&Virgo E Cl.	   &	       &  0.14   & \\ 
     233&    70216&     1868 &    7843&     4607&        -&124112.41& 115311.9&SBb? sp   		& 9.58& 3.95&	 2255&Virgo E Cl.	   &	       &  0.14   & \\ 
     234&    70214&     1869 &    7842&     4608&        -&124113.29& 100920.9&SB(r)0   		& 8.16& 4.30&	 1864&Virgo E Cl.	   &	       &  0.07   & \\ 
     235&    42205&     1883 &    7850&     4612&        -&124132.76& 071853.2&(R)SAB0   		& 8.56& 2.16&	 1875&Virgo S Cl.	   &	       &  0.11   & \\ 
     236&    70223&     1903 &    7858&     4621&        -&124202.32& 113848.9&E5  			& 6.75& 7.67&	  444&Virgo E Cl.	   &	       &  0.14   &M 59 \\ 
     237&    42208&     1923 &    7871&     4630&        -&124231.15& 035737.3&IB(s)m?   		& 9.89& 2.31&	  742&Virgo S Cl.	   &	       &  0.13   & \\ 
     238&    14109&        - &    7869&     4629&        -&124232.67&-012102.4&SAB(s)m pec 		&11.84& 1.38&	 1116&Virgo Out.	   &	       &  0.16   & \\ 
     239&    99112&     1932 &    7875&     4634&        -&124240.96& 141745.0&SBcd: sp   		& 9.25& 2.92&	  116&Virgo E Cl.	   &CP         &  0.12   &KPG351B \\ 
     240&    70229&     1938 &    7880&     4638&        -&124247.43& 112632.9&S0-   			& 8.21& 2.01&	 1147&Virgo E Cl.	   &	       &  0.11   & \\ 
     241&    43002&     1939 &    7878&     4636&        -&124249.87& 024116.0&E/S0/1;LINER;Sy3   	& 6.42& 9.63&	 1094&Virgo S Cl.	   &	       &  0.12   & \\ 
     242&    70230&     1943 &    7884&     4639&        -&124252.37& 131526.9&SAB(rs)bc;Sy1.8   	& 8.81& 3.20&	 1048&Virgo E Cl.	   &	       &  0.11   & \\ 
     243&    15008&        - &    7895&     4643&        -&124320.14& 015842.1&SB(rs)0/a;LINER   	& 7.41& 3.00&	 1346&Virgo Out.	   &	       &  0.13   & \\ 
     244&    71015&     1972 &    7896&     4647&        -&124332.45& 113457.4&SAB(rs)c   		& 8.05& 2.60&	 1422&Virgo E Cl.	   &CP         &  0.11   &KPG353A \\ 
     245&    71016&     1978 &    7898&     4649&        -&124340.01& 113309.4&E2   			& 5.74& 5.10&	 1095&Virgo E Cl.	   &CP         &  0.11   &M 60, KPG353B \\ 
     246&   100004&        - &    7901&     4651&        -&124342.63& 162336.2&SA(rs)c;LINER   		& 8.03& 3.90&	  797&Virgo Out.	   &	       &  0.12   & \\ 
     247&    71019&     1987 &    7902&     4654&        -&124356.58& 130736.0&SAB(rs)cd;HII   		& 7.74& 4.99&	 1039&Virgo E Cl.	   &	       &  0.11   & \\ 
     248&    71023&     2000 &    7914&     4660&        -&124431.97& 111125.9&E5   			& 8.21& 1.89&	 1115&Virgo E Cl.	   &	       &  0.14   & \\ 
     249&    71026&     2006 &    7920&        -&     3718&124445.99& 122105.2&S   			&11.91& 2.60&	  844&Virgo E Cl.	   &	       &  0.13   & \\ 
     250&    43018&        - &    7924&     4665&        -&124505.96& 030320.5&SB(s)0/a   		& 7.43& 4.50&	  785&Virgo Out.	   &	       &  0.11   & \\ 
     251&    15015&        - &    7926&     4666&        -&124508.59&-002742.8&SABc:;HII;LINER   	& 7.06& 4.57&	 1513&Virgo Out.	   &CP         &  0.11   & \\ 
     252&    15016&        - &    7931&     4668&        -&124532.14&-003205.0&SB(s)d:;NLAGN   		&10.58& 1.38&	 1619&Virgo Out.	   &CP         &  0.11   & \\ 
     253&    15019&        - &    7951&     4684&        -&124717.52&-024338.6&SB(r)0+;HII   		& 8.39& 2.88&	 1490&Virgo Out.	   &	       &  0.12   & \\ 
     254&    71043&     2058 &    7965&     4689&        -&124745.56& 134546.1&SA(rs)bc   		& 7.96& 5.86&	 1620&Virgo E Cl.	   &	       &  0.10   & \\ 
     255&    43028&        - &    7961&     4688&        -&124746.46& 042009.9&SB(s)cd   		&11.16& 4.40&	  984&Virgo Out.	   &	       &  0.13   & \\ 
     256&    15023&        - &       -&     4691&        -&124813.63&-031957.8&(R)SB(s)0/a pec;HII   	& 8.54& 2.82&	 1119&Virgo Out.	   &	       &  0.12   & \\ 
     257&    71045&     2070 &    7970&     4698&        -&124822.92& 082914.3&SA(s)ab;Sy2   		& 7.56& 5.67&	 1008&Virgo E Cl.	   &	       &  0.11   & \\ 
     258&        -&        - &       -&     4697&        -&124835.91&-054803.1&E6;AGN   		& 6.37& 7.24&	 1241&Virgo Out.	   &	       &  0.13   & \\ 
     259&    43034&        - &    7975&     4701&        -&124911.56& 032319.4&SA(s)cd   		& 9.77& 3.60&	  727&Virgo Out.	   &	       &  0.13   & \\ 
     260&   100011&        - &    7980&     4710&        -&124938.96& 150955.8&SA(r)0+? sp;HII   	& 7.57& 4.30&	 1129&Virgo Out.	   &	       &  0.13   & \\ 
     261&    43040&        - &    7982&        -&        -&124950.19& 025110.4&Sd(f)   			&10.17& 3.39&	 1158&Virgo Out.	   &	       &  0.15   & \\ 
     262&    43041&        - &    7985&     4713&        -&124957.87& 051841.1&SAB(rs)d;LINER   	& 9.75& 3.20&	  652&Virgo Out.	   &	       &  0.12   & \\ 
     263&   129027&        - &    7989&     4725&        -&125026.61& 253002.7&SAB(r)ab pec;Sy2   	& 6.17& 9.66&	 1209&Coma I Cl.	   &   3       &  0.05   & \\ 
     264&    15027&        - &    7991&        -&        -&125038.96& 012752.3&Sd(f) 			&11.61& 1.70&	 1272&Virgo Out.	   &	       &  0.11   & \\ 
     265&        -&        - &       -&     4720&        -&125042.78&-040921.0&Pec   			&10.77& 0.65&	 1504&Virgo Out.	   &	       &  0.11   & \\ 
     266&        -&        - &       -&     4731&        -&125101.09&-062335.0&SB(s)cd   		& 9.79& 6.61&	 1491&Virgo Out.	   &	       &  0.14   & \\ 
     267&   129028&        - &    8005&     4747&        -&125145.96& 254638.3&SBcd? pec sp   		&10.29& 3.95&	 1179&Coma I Cl.	   &   3       &  0.04   & \\ 
     268&    71060&        - &    8007&     4746&        -&125155.37& 120458.9&Sb: sp   		& 9.50& 2.20&	 1779&Virgo Out.	   &	       &  0.15   & \\ 
     269&    71062&     2092 &    8010&     4754&        -&125217.56& 111849.2&SB(r)0-:   		& 7.41& 5.03&	 1377&Virgo E Cl.	   &P	       &  0.14   &KPG356A \\ 
     270&    15029&        - &    8009&     4753&        -&125222.11&-011158.9&I0   			& 6.72& 6.03&	 1239&Virgo Out.	   &	       &  0.15   & \\ 
\noalign{\smallskip}
\hline
\end{tabular}
\]
\end{table*}
\addtocounter{table}{-1}

\begin{table*}
\tiny
\caption{continue}
\label{Tabmod}
\[
\begin{tabular}{p{0.2cm}r p{0.2cm}p{0.2cm}p{0.2cm}p{0.2cm}cc p{2.6cm}rrrc p{0.2cm}p{0.2cm}c}
\hline
\noalign{\smallskip}
     HRS&    CGCG &      VCC &    UGC &      NGC&    IC   & RA(2000) &   dec         & type  &$K_S$ $_{tot}$& $D(25)$ &    vel     &   memb  & group & AB  & Other name\\
	&	  &	     &	      &         &         & h m s    & $^{\circ}$ ' "&       &mag           & '       &  km s$^{-1}$&         &       & mag &   \\
\noalign{\smallskip}    
\hline
\noalign{\smallskip}     														      
     271&   100015&        - &    8014&     4758&        -&125244.04& 155055.9&Im:;HII   		&10.93& 3.00&	 1240& Virgo Out.	    &	       &  0.11  &  \\ 
     272&    71065&     2095 &    8016&     4762&        -&125256.05& 111350.9&SB(r)0 sp;LINER   	& 7.30& 8.70&	  985& Virgo E Cl.	    &  P       &  0.09  & KPG356B \\ 
     273&    15031&        - &    8020&     4771&        -&125321.27& 011609.0&SAd? sp;NLAGN   		& 9.01& 4.00&	 1119& Virgo Out.	    &	       &  0.09  &  \\ 
     274&    15032&        - &    8021&     4772&        -&125329.17& 021006.0&SA(s)a;LINER;Sy3   	& 8.36& 2.90&	 1038& Virgo Out.	    &	       &  0.12  &  \\ 
     275&        -&        - &       -&     4775&        -&125345.70&-063719.8&SA(s)d   		& 9.22& 2.14&	 1566& Virgo Out.	    &	       &  0.15  &  \\ 
     276&    71068&        - &    8022&     4779&        -&125350.86& 094235.7&SB(rs)bc;Sbrst   	& 9.87& 2.10&	 2832& Virgo Out.	    &	       &  0.09  &  \\ 
     277&    43060&        - &       -&     4791&        -&125443.97& 080310.7&cI   			&11.35& 1.20&	 2529& Virgo Out.	    &	       &  0.14  &  \\ 
     278&    71071&        - &    8032&        -&        -&125444.19& 131414.2&S   			&10.39& 2.75&	 1121& Virgo Out.	    &	       &  0.15  &  \\ 
     279&    15037&        - &    8041&        -&        -&125512.68& 000700.0&SB(s)d   		&11.98& 3.10&	 1321& Virgo Out.	    &	       &  0.10  &  \\ 
     280&    43066&        - &    8043&     4799&        -&125515.53& 025347.9&S?   			& 9.89& 1.60&	 2802& Virgo Out.	    &	       &  0.15  &  \\ 
     281&    43068&        - &    8045&        -&        -&125523.62& 075434.0&IBm:    			&11.82& 0.91&	 2801& Virgo Out.	    &	       &  0.18  &  \\ 
     282&    43069&        - &       -&     4803&        -&125533.67& 081425.8&Comp   			&10.71& 0.50&	 2664& Virgo Out.	    &	       &  0.13  &  \\ 
     283&    43071&        - &    8054&     4808&        -&125548.94& 041814.7&SA(s)cd:;HII   		& 9.04& 2.60&	  760& Virgo Out.	    &	       &  0.16  &  \\ 
     284&        -&        - &       -&        -&     3908&125640.62&-073346.1&SB(s)d?;HII   		& 9.10& 1.82&	 1296& Virgo Out.	    &	       &  0.15  &  \\ 
     285&    15049&        - &    8078&     4845&        -&125801.19& 013433.0&SA(s)ab sp;HII   	& 7.79& 5.20&	 1097& Virgo Out.	    &	       &  0.09  &  \\ 
     286&    71092&        - &    8102&     4866&        -&125927.14& 141015.8&SA(r)0+: sp;LINER   	& 7.92& 6.00&	 1986& Virgo Out.	    &	       &  0.12  &  \\ 
     287&    15055&        - &    8121&     4904&        -&130058.67&-000138.8&SB(s)cd;Sbrst   		& 9.50& 2.40&	 1174& Virgo Out.	    &	       &  0.11  &  \\ 
     288&        -&        - &       -&     4941&        -&130413.14&-053305.8&(R)SAB(r)ab:;Sy2   	& 8.22& 3.63&	 1114& Virgo Out.	    &	       &  0.16  &  \\ 
     289&        -&        - &       -&     4981&        -&130848.74&-064639.1&SAB(r)bc;LINER   	& 8.49& 2.75&	 1678& Virgo Out.	    &	       &  0.18  &  \\ 
     290&   189037&        - &    8271&     5014&        -&131131.16& 361654.9&Sa? sp   		&10.11& 1.70&	 1136& Canes Ven. Spur      & 3        &  0.03  &  \\ 
     291&   217031&        - &    8388&     5103&        -&132030.08& 430502.3&Sab    			& 9.49& 1.45&	 1297& Canes Ven. Spur      &	       &  0.08  &  \\ 
     292&   218010&        - &    8439&     5145&        -&132513.92& 431602.2&S?;HII;Sbrst   		& 9.33& 2.00&	 1225& Canes Ven. Spur      &	       &  0.05  &  \\ 
     293&    16069&        - &    8443&     5147&        -&132619.71& 020602.7&SB(s)dm   		& 9.73& 1.91&	 1092& Virgo Out.	    &	       &  0.12  &  \\ 
     294&   246017&        - &    8593&        -&      902&133601.22& 495739.0&Sb    			&10.42& 2.19&	 1608& Canes Ven. Spur      &	       &  0.05  &  \\ 
     295&    73054&        - &    8616&     5248&        -&133732.07& 085306.2&(R)SB(rs)bc;Sy2;HII     	& 7.25& 1.79&	 1152& Virgo-Libra Cl.      & P        &  0.11  &  \\ 
     296&   190041&        - &    8675&     5273&        -&134208.34& 353915.2&SA(s)0;Sy1.9   		& 8.67& 2.75&	 1064& Canes Ven. Spur      &	       &  0.04  & KPG391A \\ 
     297&   246023&        - &    8711&     5301&        -&134624.61& 460626.7&SA(s)bc: sp   		& 9.11& 4.17&	 1508& Canes Ven. Spur      &	       &  0.07  &  \\ 
     298&   218047&        - &    8725&     5303&        -&134744.97& 381816.4&Pec   			&10.23& 0.91&	 1419& Canes Ven. Spur      & CP       &  0.06  & KPG397A \\ 
     299&    45108&        - &    8727&     5300&        -&134816.04& 035703.1&SAB(r)c    		& 9.50& 3.89&	 1171& Virgo-Libra Cl.      &	       &  0.10  &  \\ 
     300&   218058&        - &    8756&        -&        -&135035.89& 423229.5&Sab    			&10.34& 1.70&	 1354& Canes Ven. Spur      &	       &  0.06  &  \\ 
     301&    17088&        - &    8790&     5334&     4338&135254.46&-010652.7&SB(rs)c:   		& 9.94& 4.17&	 1380& Virgo-Libra Cl.      &	       &  0.20  &  \\ 
     302&    45137&        - &    8821&     5348&        -&135411.27& 051338.8&SBbc: sp 		&10.87& 3.55&	 1443& Virgo-Libra Cl.      &	5      &  0.12  &  \\ 
     303&   295024&        - &    8843&     5372&        -&135446.01& 583959.4&S?   			&10.65& 0.65&	 1717& Canes Ven-Came. Cl.  &	       &  0.04  &  \\ 
     304&    46001&        - &    8831&     5356&        -&135458.46& 052001.4&SABbc: sp;HII 		& 9.64& 3.09&	 1370& Virgo-Libra Cl.      &	5      &  0.11  &  \\ 
     305&    46003&        - &    8838&     5360&      958&135538.75& 045906.2&I0   			&11.15& 2.19&	 1171& Virgo-Libra Cl.      &	5      &  0.13  &  \\ 
     306&    46007&        - &    8847&     5363&        -&135607.21& 051517.2&I0?   			& 6.93& 4.07&	 1136& Virgo-Libra Cl.      &	5      &  0.12  &  \\ 
     307&    46009&        - &    8853&     5364&        -&135612.00& 050052.1&SA(rs)bc pec;HII   	& 7.80& 6.76&	 1242& Virgo-Libra Cl.      &	5      &  0.12  &  \\ 
     308&    46011&        - &    8857&        -&        -&135626.61& 042348.0&Sb(f)   			&11.93& 0.91&	 1091& Virgo-Libra Cl.      &	       &  0.14  &  \\ 
     309&   272031&        - &    9036&     5486&        -&140724.97& 550611.1&SA(s)m: 			&11.95& 1.86&	 1383& Canes Ven-Came. Cl.  & 3        &  0.09  &  \\ 
     310&    47010&        - &    9172&     5560&        -&142005.42& 035928.4&SB(s)b pec   		& 9.98& 3.72&	 1718& Virgo-Libra Cl.      &	3      &  0.13  & ARP286 \\ 
     311&    47012&        - &    9175&     5566&        -&142019.95& 035600.9&SB(r)ab;LINER   		& 7.39& 6.61&	 1492& Virgo-Libra Cl.      &	3      &  0.13  & ARP286 \\ 
     312&    47020&        - &    9183&     5576&        -&142103.68& 031615.6&E3   			& 7.83& 3.55&	 1482& Virgo-Libra Cl.      &	3      &  0.14  &  \\ 
     313&    47022&        - &    9187&     5577&        -&142113.11& 032608.8&SA(rs)bc:   		& 9.75& 3.39&	 1490& Virgo-Libra Cl.      &	3      &  0.18  &  \\ 
     314&    19012&        - &    9215&        -&        -&142327.12& 014334.7&SB(s)d 			&10.54& 2.19&	 1389& Virgo-Libra Cl.      &	       &  0.14  &  \\ 
     315&   220015&        - &    9242&        -&        -&142521.02& 393222.5&Sc   			&11.73& 5.01&	 1440& Canes Ven. Spur      &	       &  0.05  &  \\ 
     316&    47063&        - &    9308&     5638&        -&142940.39& 031400.2&E1   			& 8.25& 2.69&	 1845& Virgo-Libra Cl.      &	4      &  0.14  &  \\ 
     317&    47066&        - &    9311&        -&     1022&143001.85& 034622.3&S?   			&11.70& 1.10&	 1716& Virgo-Libra Cl.      &	       &  0.15  &  \\ 
     318&    47070&        - &    9328&     5645&        -&143039.35& 071630.3&SB(s)d   		& 9.69& 2.40&	 1370& Virgo-Libra Cl.      &	       &  0.12  &  \\ 
     319&    75064&        - &    9353&     5669&        -&143243.88& 095330.5&SAB(rs)cd   		&10.35& 3.98&	 1368& Virgo-Libra Cl.      &	       &  0.12  &  \\ 
     320&    47090&        - &    9363&     5668&        -&143324.34& 042701.6&SA(s)d   		&11.71& 3.31&	 1583& Virgo-Libra Cl.      & P        &  0.16  &  \\ 
     321&    47123&        - &    9427&     5692&        -&143818.12& 032437.2&S?   			&10.54& 0.89&	 1581& Virgo-Libra Cl.      &	       &  0.16  &  \\ 
     322&    47127&        - &    9436&     5701&        -&143911.06& 052148.8&(R)SB(rs)0/a;LINER    	& 8.14& 4.27&	 1505& Virgo-Libra Cl.      &	       &  0.16  &  \\ 
     323&    48004&        - &    9483&        -&     1048&144257.88& 045324.5&S   			& 9.55& 2.24&	 1640& Virgo-Libra Cl.      & CP       &  0.16  &  \\ 
\noalign{\smallskip}
\hline
\end{tabular}
\]
\end{table*}

\clearpage

\section{SPIRE observations}

With its large field of view (4'x8'), its sensitivity ($S$ $\sim$ 7 mJy, 5 
$\sigma$ in 1 hour, point source mode), and angular resolution 
($\sim$ 30 arcsec; see Table \ref{TabSPIRE}), SPIRE 
on Herschel is the ideal instrument for the proposed survey.

\begin{table}
\caption{The SPIRE performances}
\label{TabSPIRE}
\[
\begin{tabular}{cccc}
\hline
\noalign{\smallskip}
$\lambda$  & 250 $\mu$m  & 350 $\mu$m  & 500 $\mu$m  \\
Ang.~Res.  & 18.1"       &  25.2"      &  36.9"      \\
Lin.~Res.  & 1.7 kpc     & 2.4 kpc     & 3.6 kpc     \\
Sensitivity& 12 mJy/beam & 8  mJy/beam & 12 mJy/beam \\
\noalign{\smallskip}
\hline
\end{tabular}
\]
Note: 1-$\sigma$ sensitivity (instrumental noise) for a standard map (one repeat = two cross-linked scans) as determined during the science verification phase
(on board observations).
The linear resolution is at 20 Mpc.
\end{table}

In constructing our observing program, we have used different integration
times for early- and late-type galaxies because the former are known to contain
less dust. 
\\

Integration times for early-types were determined by
assuming a lower limit of 
$\sim$ 10$^4$ M$\odot$ to their total dust mass as determined from IRAS observations 
(Bregman et al. 1998) or measured from optical absorption line measurements 
(Goudfrooij et al. 1994; Van Dokkum \& Franx 1995; Ferrari et al. 1999). At 20 Mpc, a galaxy with 
a dust mass of $\sim$ 10$^4$ M$\odot$ would have a flux density at 250 $\mu$m of 11 mJy. 
We estimate that with the adopted integration times we should detect 
an elliptical with $\sim$ 10$^4$ M$\odot$ of dust at the 3$\sigma$ level.\\

Integration times for spirals have been chosen so that we should be able to
detect dust outside the optical radius\footnote{Defined as the B band isophotal radius at 25 mag arcsec$^{-2}$}, for which there is evidence from ISO
observations (Alton et al. 1998; Davies et al. 1999).
By combining ISOPHOT (Alton et al. 1998)
and SCUBA (Vlahakis et al. 2005) observations of extended sources 
with the spectral energy distribution of normal galaxies of different type (Boselli 
et al. 2003), and assuming a standard gas to dust ratio of $\sim$ 100 with the extragalactic calibration for
the dust-emissivity coefficient (James et al. 2002), we estimate that 
the dust associated with the extended HI disc would have a flux density of $\sim$ 14 mJy
per beam at 250 $\mu$m for HI column densities of $\sim$ 10$^{20}$ atoms cm$^{-2}$.
In normal galaxies such gas column densities are generally reached well outside 
the stellar disc, at $\sim$ 1.5 optical radii (Cayatte et al. 1994; Broelis \& Van Woerden 1994).
Because of metallicity gradients, however, the gas to dust ratio is expected to increase in 
the outer discs, as observed in M31 by Cuillandre et al. (2001). We would therefore expect to observe lower far-IR flux densities.
With 12 minutes of integration time (3 scan-maps) we will 
get to an instrumental noise of 6.9 mJy/beam, when the expected 
emission is 14 mJy per beam; this sensitivity will allow us to detect the dust emission in the outer disc
by integrating flux densities along elliptical annuli with a sensitivity 
of $\sim$ 30 better than that of previous observations (Alton et al. 1998).\\

All galaxies outside the Virgo cluster will be observed in scan-map mode (30"/sec). 
For each galaxy, the size of the scan-map has been chosen based on the optical size
of the galaxy. For late-type galaxies, the observing mode has been chosen so that
our images will cover the galaxy at least out to 1.5 $\times$ $D_{25}$ (where $D_{25}$ is the 25 mag arcsec$^{-2}$ isophotal diameter)
i.e. out to the approximate HI radius.
For spirals we will make three pairs of cross-linked observations, which will reach a 1 $\sigma$ sensitivity in 
instrumental noise of 6.9, 4.6 and 6.9 mJy/beam at 250, 350 and 500 $\mu$m respectively
as determined during the science verification phase. These values are comparable to the noise
from the confusion of faint unrelated sources (4, 5 and 6 mJy/beam in the three bands).
Elliptical galaxies are likely to be weaker and thus need longer integration times, we have compromised by only requiring
the image to contain the galaxy out to $D_{25}$. 
For ellipticals and S0s we will make eight pairs of cross-linked observations
reaching a 1$\sigma$ sensitivity in instrumental noise
of 4.2, 2.8 and 4.2 mJy/beam at 250, 350 and 500 $\mu$m respectively, which in
the latter two bands is below the noise expected from the confusion of faint sources.\\

Sixty square degrees centered on the Virgo cluster region will be covered by Herschel in parallel mode (i.e. using both SPIRE and 
PACS) as part of the Herschel Virgo Cluster Survey HeViCS \footnote{www.arcetri.astro.it/hevics}. Eighty-three HRS galaxies fall into this region. 
To avoid duplication, these galaxies will be observed only during the HeViCS survey. 
HeViCS will survey this region in fast-scan mode (60"/sec) using both SPIRE and PACS. The survey will make eight scans 
reaching a 1 $\sigma$ of instrumental noise of 4.5, 6.2 and 5.3 mJy/beam at 250, 350 and 500 $\mu$m respectively, 
as determined using the Herschel Observation Planning Tool (HSpot v4.2.0).
This is not as deep as our observations of ellipticals and S0s outside Virgo, but the effect of confusion noise means that 
the decrease in effective sensitivity is really small. On the plus side, for the HRS galaxies in Virgo we will also have PACS images at 110 and 160 $\mu$m.



\section{Complementary data}

A number of key aims of our survey (see Section 2) can only be achieved through the combination
of Herschel observations with corollary data. 
UV to near-IR imaging is needed to
determine the distribution and content of the different stellar populations, 
to quantify the intensity of the interstellar radiation field and to
study the recent activity of star formation (UV and H$\alpha$). Optical
spectroscopy is crucial to measure stellar and gas metallicities and
to determine the presence of any nuclear activity. At the same time, 
the Balmer emission line can be used to measure the dust extinction within HII regions 
and are thus essential for quantifying the current level of star formation activity. 
HI and CO line data are necessary for determining the content and distribution of
the gaseous component of the ISM, the principal feeder of star formation in galaxies. 
High resolution HI imaging can also be used to study the kinematical properties of the target
galaxies.
Mid-, far-IR and sub-mm data, combined with SPIRE data, will be used to study the physical
properties of the different dust components (PAHs, very small grains and big grains), 
and radio continuum data for measuring the thermal and synchrotron emission.\\

Given its definition, the HRS is easily accessible for ground based and
space facilities: the selected galaxies are in fact relatively bright ($m_B$ $<$
 15 mag) and extended ($\sim$ 2-3 arcmin). Listed below are the most
 important references for ancillary data.

\subsection{UV, optical, near-IR and H$\alpha$ imaging}
Of the 323 galaxies in the HRS, 280 have been observed  
by the Galaxy Evolution Explorer (GALEX) in the two UV bands FUV ($\rm
\lambda_{eff}=1539\AA, \Delta \lambda=442\AA$) and NUV ($\rm
\lambda_{eff}=2316\AA, \Delta \lambda=1069\AA$). Observations have been 
taken as part of the Nearby Galaxy Survey (NGS; Gil de Paz et al. 2007), 
the Virgo cluster survey (Boselli et al. 2005), the All Imaging Survey (AIS) or as pointed observations in
open time proposals. A GALEX legacy survey has been recently accepted to complete the UV observation
of the HRS galaxies at a deepness of the Medium Imaging Survey (1500 sec/field).\\

SDSS (DR7 release, Adelman-McCarthy et al. 2008) images in the 
$ugriz$ filters are available for 313  objects. All galaxies have been 
observed in the JHK filters by 2MASS (Jarrett et al. 2003). Deep B,V, H and K
band images for all Virgo cluster galaxies and for some Coma I Cloud galaxies are
available on the GOLDMine database (http://goldmine.mib.infn.it/; Gavazzi et al. 2003).
These have been taken with pointed observations during the near infrared and
optical extensive surveys of the Virgo cluster carried out by Boselli et al. (1997, 2000; 2003) and Gavazzi et al. (2000). \\

An H$\alpha$+[NII] imaging survey of the star forming HRS galaxies
found outside the Virgo cluster under way at the 2.1m San Pedro 
Martir telescope is almost complete. Combined with images taken 
in the Virgo cluster region (Boselli \& Gavazzi 2002, Boselli et al. 2002 and Gavazzi et al. 1998; 2002; 2006) 
H$\alpha$+[NII] data are now available for 221 of the 258 late-type objects and 26 of the 65 early-types.

\subsection{Integrated spectroscopy}

A low resolution ($R$ $\sim$ 1000), integrated spectroscopic survey in the wavelength range
3500-7200 \AA~ is under way at the 1.93m OHP telescope.
In order to sample the
spectral properties of the whole galaxy, and not just those of
the nuclear regions (these last provided by the SDSS for 106 galaxies in the DR6), observations have been
executed using the drifting technique described in Kennicutt (1992). 
Exposures
are taken while constantly and uniformly drifting the spectrograph slit
over the full extent of the galaxy. 
A resolution $R$ $\sim$ 1000
is mandatory for resolving [NII] from H$\alpha$
and measuring the stellar underling Balmer absorption under H$\beta$.\\

Data for 64 Virgo cluster galaxies 
in the HRS have already been published in Gavazzi et al. (2004) along with a few other galaxies 
in Moustakas \& Kennicutt (2006), Jansen et al. (2000) and Kennicutt (1992b).
We have data from the literature or from our own observations  for 256/258 of the late-types and 33/65 of the early-types, 
making the whole sample complete at 89 \%. The remaining objects will be included in the future observing runs.

\subsection{HI and CO lines}

Single beam HI observations are available from a large variety of sources. 
Most of the galaxies have HI data in Springob et al. (2005) and 
Gavazzi et al. (2005), this last reference limited to the Virgo cluster region.
Out of the 8 late-type galaxies, 249 have an HI measurements, as do 55/65 of the early-types.
All galaxies in the 0$^\circ$ $<$ dec $<$ 30$^\circ$ range (80 \%) will be observed during the
ALFALFA survey under way at the 305m Arecibo radio telescope (Giovanelli et al. 2005).
Data will be gathered with an homogeneous sensitivity (rms=2.5 mJy) 
and spectral resolution (5.5 km s$^{-1}$): at the average distance of the HRS the ALFALFA 
survey will detect all sources with $MHI$ $\geq$ 10$^{7.5}$ M$\odot$.
VLA and WSRT HI maps are available for 236 HRS galaxies, from the WHISP (van der Hulst 2002) and VIVA (Chung et al. 2009) 
survey, this last limited to the Virgo region, from VLA archives or from our own WSRT observations.\\

A $^{12}$CO(1-0) line (2.6 mm, 115 GHz) survey of the HRS galaxies is under way at the 12m Kitt Peak
telescope. Fifty-eight spiral galaxies have been observed to date, with an average rms of 3 mK in T$_R^*$ scale.
Thanks to these new observations and to data in the literature CO 
measurements are now available for 161/258 late-type and 22/65
early-type galaxies with a detection rate of 83\% and 55\% respectively. Data have been taken 
from different sources, mostly from the FCRAO survey of Young et al. (1995), 
or Kenney \& Young (1988) and Boselli et al. (1995; 2002) for the Virgo cluster region.
We obtained time at the James Clerk Maxwell Telescope (JCMT) on the 345 GHz HARP array receiver 
to search for CO(3-2) emission from the central 3' x 3' region (with 15 arcsec resolution) of a subsample 
of late type galaxies with $K_{Stot}<8.7$. As of the beginning of January 2010, 42 of a total of 56 galaxies 
in the subsample have been observed to completion.

\subsection{Mid-IR, far-IR and sub-mm}

Infrared data for the HRS galaxies are already available from different sources. 208/258
late-type and 32/65 of the early-type galaxies have been detected by IRAS at 60 and 100 $\mu$m,
only 103 (Sa-Sd-Im-BCD) and 17 (E-S0-S0a) at 12 and 25 $\mu$m. 
Virgo galaxies have been observed with
CAM (45), PHOT (26) and LWS (21) on ISO (Leech et al. 1999; Malhotra et al. 2001; Roussel et al. 2001; Boselli et al. 2002, 2003; 
Tuffs et al. 2002). 
Spitzer
observations have been completed for 157 HRS galaxies with IRAC and 181
galaxies with MIPS. The data for all of these galaxies will eventually be
available from the Spitzer archives. The pipeline-processed IRAC
data is sufficient for science analysis, but MIPS data has been
reprocessed using the MIPS Data Analysis Tools (Gordon et al. 2005)
along with additional processing software.
\\

All HRS galaxies have been observed during the recently completed AKARI all sky survey in the 9, 20, 70, 90
and 160 $\mu$m bands which covered $\sim$ 95\% of the whole sky with an angular 
resolution comparable to that of SPIRE ($\sim$ 30 arcsec at 90 $\mu$m) and a sensitivity $\sim$ three times better than IRAS. Pointed observations
with AKARI and SCUBA are also available of a few 
ellipticals.

\subsection{Radio continuum data}

The HRS galaxies have been observed by the NVSS survey at 20 cm (1415 MHz) (Condon et al. 1998) 
with an angular resolution of 45 arcsec and a sensitivity of 2.5 mJy. 22/65 and 159/258 
of the early- and late-type galaxies respectively have been detected at 20 cm. The data of the FIRST survey, at 5 arcsec
resolution, are also available (Becker et al. 1995). For a fraction of galaxies data at 6.3 and 2.8 cm 
are also available 
from Niklas et al. (1995) and Vollmer et al. (2004).

\subsection{X-ray data}

X-ray data at 0.2-4 keV from the Einstein Observatory imaging instruments (IPC and HRI) 
are available for 31 early-type and 52 late-type galaxies from Fabbiano et al. (1992) and Shapley et al. (2001).
Ten HRS galaxies have been detected by the ROSAT all sky survey (Voges et al. 1999), while Chandra observations of
many of the early-type objects within the HRS are still under way (Gallo et al. 2008).\\

Table \ref{Tabdata} shows the availability of ancillary data for the HRS galaxies.

\section{HRS data processing, data products and WEB site}

The HRS galaxies are distributed widely across the sky, the survey will progress along with
the Herschel mission, that is expected to last for three and a half years, with the first six months
for commissioning, performance verification, science demonstration and the last three years for science. 
Data for each single observation will be proprietary for one year after
it is taken.
During this time, the data will be processed using the most up-to-date
pipeline developed by the SPIRE-Instrument Control Centre since some members are also part of
our consortium. This ensures we have access to the latest calibration
files and bug fixes.
Aside from the default data processing, the pipeline will also
include new steps developed by the key-project teams (such as cosmic ray removal and
baseline subtraction) for producing the most accurate representation of large-scale
structures, such as the extended emission around galaxies. Additional post-processing steps will be applied
to prepare the data so that it can be easily used by the broader
astronomical community. This will include reformatting the headers of the fits files so that they
are easy to understand. Furthermore, the data will have been visually
inspected so as to ensure that they suffer from no severe artefacts
related to the observations or data reduction, and in cases where
problems are identified, the data will be manually re-processed to
obtain the best results.
\\
After the proprietary time the team plans to make the reduced data 
available to the community through a dedicated web page.
Herschel and ancillary data will be archived in an Information System at the Laboratoire d'Astrophysique de Marseille 
developped under the SITools middleware interface (http://vds.cnes.fr/sitools/) with a Postgresql database.
Images will be stored as fits files, and distributed in fits format while catalogues will be distributed in VO table and ASCII (cvs) format.
The SPIRE data delivered by the team will be optimized for use by the
wider astronomical community.  
\\
Given the legacy nature of this project, we will also present through an interactive database
all the ancillary data, X, UV, optical, near- and far-IR, sub-mm and radio continuum fluxes and images,
optical and line spectroscopy and derived parameters (structural parameters, metallicities, SFR...)
with the aim of providing the community with a unique dataset, a suitable reference for future studies.

\section{The statistical properties of the Herschel Reference Sample}

In this section, we investigate how representative the HRS is of the galaxy population as a whole, thus unbiased versus cosmic variance. This is necessary
since, as selected, the HRS is biased versus galaxies located in a high density environment (Virgo). One way to test this 
is to see whether the luminosity distributions at different wavelengths in the HRS are similar to the local galaxy luminosity functions at the corresponding
frequencies. 
\\

\begin{table}
\caption{The completeness in the data sample}
\label{Tabcompl}
\[
\begin{tabular}{ccc}
\hline
\noalign{\smallskip}
$\lambda$  		& E+S0+S0a	& Sa-Sm-Im-BCD	\\
K  			& 	100\%	& 100\%		\\
HI  			& 85\% (40\%)	& 97\% (93\%)	\\
FIR~(60 $\mu$m) 	& 95\% (62\%)	& 89\% (83\%)	\\
Radio~cont.~(20 cm)	& 100\% (34\%)	& 100\% (62\%)	\\
\noalign{\smallskip}
\hline
\end{tabular}
\]
Note: Values in parenthesis give the detection rate.
\end{table}

\begin{figure}
\epsscale{1.5} 
\plotone{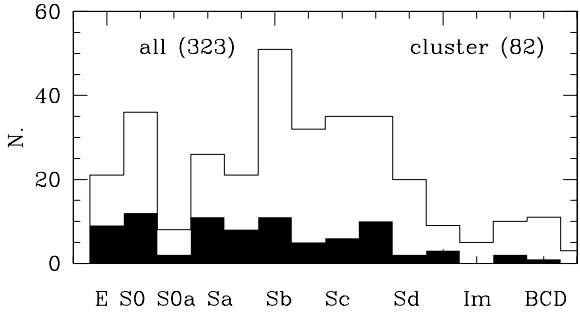}
\small{\caption{The distribution in morphological type of the HRS for all (empty histogram) 
and cluster (filled histogram) galaxies. The cluster sample is composed of galaxies
members of the Virgo A and B clouds.
\label{stat}}}
\end{figure}

\begin{figure}
\epsscale{1.5} 
\plotone{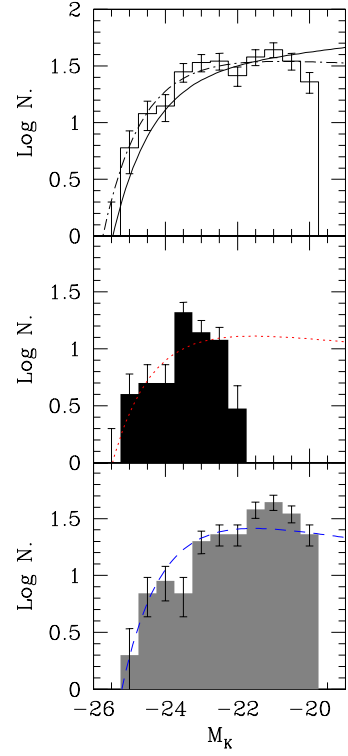}
\small{\caption{The K band luminosity distribution of the HRS for all galaxies 
is compared to the 2MASS K band luminosity function of Kochanek 
et al. (2001) (black solid line)
and Cole et al. (2001) (black dotted-dashed line)(upper panel). The K band luminosity distributions
of the E-S0-S0a (central panel) and Sa-Sd-Im-BCD (lower panel) galaxies in the HRS are compared to  
the Kochanek et al. (2001) K band luminosity function of early-type (red dotted line) and 
late-type (blue dashed line) galaxies. Poisson errors are also indicated.
\label{statK}}}
\end{figure}

Shown in Fig. \ref{statK} is a comparison of
the HRS luminosity distribution in the K-band (histogram) to the K band luminosity function 
of Cole et al. (2001) (dotted-dashed line; $\alpha$ = 0.96 $\pm$ 0.05, $M_K^*$ = -23.44 $\pm$ 0.03) and 
Kochanek et al. (2001) (solid line; $\alpha$ = 1.09 $\pm$ 0.06, $M_K^*$ = -23.39 $\pm$ 0.05), this last also 
considering separately early- (red dotted line; $\alpha$ = 0.92 $\pm$ 0.10, $M_K^*$ = -23.53 $\pm$ 0.06) 
and late-type (blue dashed line; $\alpha$ = 0.87 $\pm$ 0.09, $M_K^*$ = -22.98 $\pm$ 0.06) galaxies.
The Kochanek et al. (2001) values of the K band luminosity function have been corrected to take into account the different 
assumed H$_0$, while the Cole et al. (2001) $M_S^*$ Kron magnitudes have been converted into total magnitudes as
in this work. The zero point of the plotted luminosity functions is determined
assuming the same number of objects as HRS within the given magnitude limit. 
Figure \ref{statK} shows that, when considered from the perspective of the K-band, the HRS is a good approximation
to a volume-limited sample down to an absolute magnitude of $\rm M_K \simeq -20$. This
absolute magnitude limit is relatively high and explains
why the HRS is under-represented in Im and BCD galaxies (Fig. \ref{stat}). The sample includes only
the brightest ellipticals and lenticulars, whose K band luminosity function ends at $M_K$ $\sim$ -18, while it does not
include any quiescent dwarf system ($M_K$ $\geq$ -20).

\begin{figure}
\epsscale{1.5} 
\plotone{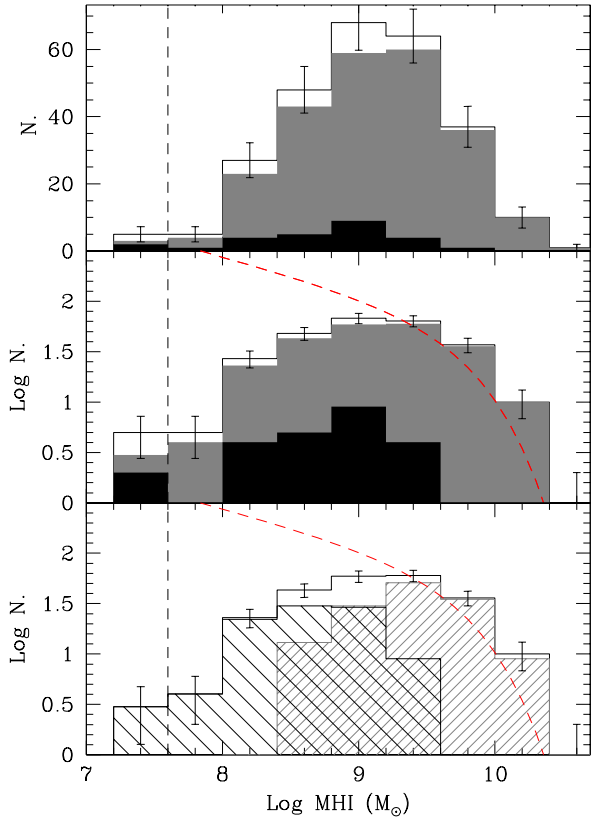}
\small{\caption{The atomic hydrogen mass distribution in linear (upper panel) and logarithmic (middle panel) scales
for the HI detected HRS galaxies (empty histogram). The black and gray histograms are for early-type (E-S0-S0a, 35\% detected) 
and late-type (Sa-Sd-Im-BCD; 93 \% detected) galaxies respectively. 
Lower panel: The atomic hydrogen mass distribution
for late-type galaxies with a normal HI gas content (HI-deficiency $<$ 0.4; tiny spaced hashed histogram)
and gas poor galaxies (HI-deficiency $\geq$ 0.4; large spaced hashed histogram).
The red dashed line
is the HI mass function determined by Zwaan et al. (2005). The vertical, dashed line indicates the detection limit of the ALFALFA
survey (2.5 mJy) at the distance of 20 Mpc.
\label{statHI}}}
\end{figure}

\begin{figure}
\epsscale{1.5} 
\plotone{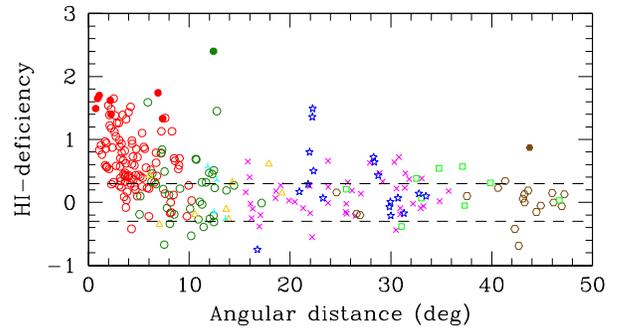}
\small{\caption{The variation of the HI-deficiency parameter as a function of the 
angular distance from the core of the Virgo cluster (M87). Red circles are for Virgo cluster galaxies
(Virgo A, B, N, S and E), dark green circles for galaxies in the Virgo outskirts,
orange triangles for the Coma I Cloud, magenta crosses for the Leo Cloud, brown hexagons for 
Ursa Major Southern Spur and Cloud, cyan three skeletal for the Crater Cloud, light green squares for Canes Venatici Spur and Camelopardalis
and blue stars for the Virgo-Lybra Cloud. Empty symbols are for HI-detected galaxies,
filled symbols for upper limits. 
\label{def}}}
\end{figure}

Figure \ref{statHI} shows the HI mass distribution of the HI detected HRS early-type (black histogram) and
late-type (gray histogram) galaxies compared to the HIPASS HI mass function determined by Zwaan et al. (2005) 
(dashed line, $\alpha$ = -1.37 $\pm$ 0.03, M$_{HI}$$^*$ = 10$^{9.8 \pm 0.03}$ M$\odot$ and $\phi$ $^*$ 
arbitrarily normalized to match the data). 
The sample includes galaxies with HI masses ranging in between $\sim$ 10$^8$ and 10$^{10}$
M$\odot$ but it is clearly under-representing galaxies with HI masses in the range 10$^{7.5}$ M$\odot$ $\leq$ M$_{HI}$ 
$\leq$ 10$^9$ M$\odot$ which are easily detectable by an HI blind survey such as ALFALFA. The HIPASS HI mass function is
dominated by gas rich, star forming galaxies, while the HRS also includes quiescent, gas-poor early-types and
HI-deficient cluster galaxies, as clearly observed in the
Virgo cluster (Gavazzi et al. 2005; 2008) or in A1367 (Cortese et al. 2008b). 
The difference between the HIPASS mass function and the HRS HI mass distribution
is due to i) the fact that the HRS is K band selected and thus does not include gas rich, low luminosity star forming systems with HI masses 
in the range 10$^{7.5}$ $\leq$ $MHI$ $\leq$ 10$^{8.5}$ M$\odot$ with K band magnitudes 12 $\leq$ K $\leq$ 16,  
ii) the presence of cluster HI-deficient late-type galaxies. Indeed the HI mass distribution of
galaxies with a normal HI gas content (HI-deficiency \footnote{The HI-deficiency parameter
is defined as as the logarithmic difference between the average HI mass of a
reference sample of isolated galaxies of similar type and linear
dimension and the HI mass actually observed in individual objects:
$HI-def$ = Log$MHI_{ref}$ - Log$MHI_{obs}$. According to Haynes \&
Giovanelli (1984), Log$MHI_{ref}$ = $a$ + $b$ $\times$ Log$(diam)$, where $a$
and $b$ are weak functions of the Hubble type and $diam$ is the linear
diameter of the galaxy (see Gavazzi et al. 2005).} $<$ 0.4) is more similar to the HIPASS mass function than the distribution 
of gas poor objects (HI-deficiency $\geq$ 0.4). 
Most of the late-type HI-deficient objects (HI-deficiency $\geq$ 0.4)
in the HRS are indeed located in the core of the Virgo cluster
as shown in Fig. \ref{def}. Galaxies in the outskirt of the Virgo cluster
or in the surrounding clouds have a normal HI gas content (see Table \ref{Tabdef})
and can be generally considered as unperturbed objects (unless belonging to groups or pairs) 
in the study of the effects of the environment on the dust properties of galaxies.
Indeed, despite the relatively poor statistics, the average HI-deficiency of late-type galaxies in the 
Coma I Cloud, Leo Cloud, Ursa Major Cloud and Southern Spur, Crater Cloud, Canes Venatici Spur, Canes Venatici - Camelopardalis Cloud   
and Virgo - Libra Cloud is consistent with that of unperturbed galaxies in the reference sample of 
Haynes \& Giovanelli (1984; HI-deficiency $\leq$ 0.3). The HI-deficiency of late-type galaxies 
in the different substructures of the Virgo cluster are generally consistent with those observed in a larger sample by Gavazzi et al. (1999a).
The only exception is the Virgo E Cloud that we found dominated by HI-deficient objects. We notice however that
the majority of the most HI-deficient galaxies (HI-deficiency $>$ 1.0) of the HRS belonging to this substructure
are classified as early-types in the VCC.\\
The Coma I Cloud, previously thought to include gas deficient objects (Garcia-Barreto et al. 1994; Gerin \& Casoli 1994),
is composed of galaxies with a normal gas content (Boselli \& Gavazzi 2009).
A detailed and complete study of the HI properties of late-type galaxies belonging to the other Clouds has not been done so far.
A rather normal HI gas content of the spiral galaxies in the Ursa Major Cluster, a substructure of the Ursa Major Cloud, has been
observed by Verheijen \& Sancisi (2001).

\begin{table*}
\caption{The average HI-deficiency of late-type galaxies in the different substructures}
\label{Tabdef}
\[
\begin{tabular}{cccc}
\hline
\noalign{\smallskip}
Substructure  				& average	& standard~deviation & N. of objects	\\
Virgo A  				&0.91 	&$\pm$ 0.47 	& 33 \\
Virgo B					&0.71 	&$\pm$ 0.46 	& 25 \\
Virgo N Cloud 				&0.39 	&$\pm$ 0.24 	& 12 \\
Virgo E Cloud				&0.83	&$\pm$ 0.53 	& 12 \\
Virgo S Cloud				&0.47	&$\pm$ 0.47	& 26 \\
Virgo Outskirts				&0.26	&$\pm$ 0.60	& 38 \\
Coma I Cloud				&0.07	&$\pm$ 0.35	& 8  \\
Leo Cloud				&0.11	&$\pm$ 0.31	& 45 \\
Ursa Major Cloud			&0.05	&$\pm$ 0.34	& 15 \\
Ursa Major Southern Spur		&-0.02	&$\pm$ 0.20	&  4 \\
Crater Cloud				&0.13	&$\pm$ 0.40	&  4 \\
Canes Ven. Spur \& Camelopardalis	&0.19	&$\pm$ 0.03	&  9 \\
Virgo - Libra Cloud 			&0.31	&$\pm$ 0.35	& 18 \\
\noalign{\smallskip}
\hline
\end{tabular}
\]
Note: Upper limits are considered as detections
\end{table*}

\begin{figure}
\epsscale{1.5} 
\plotone{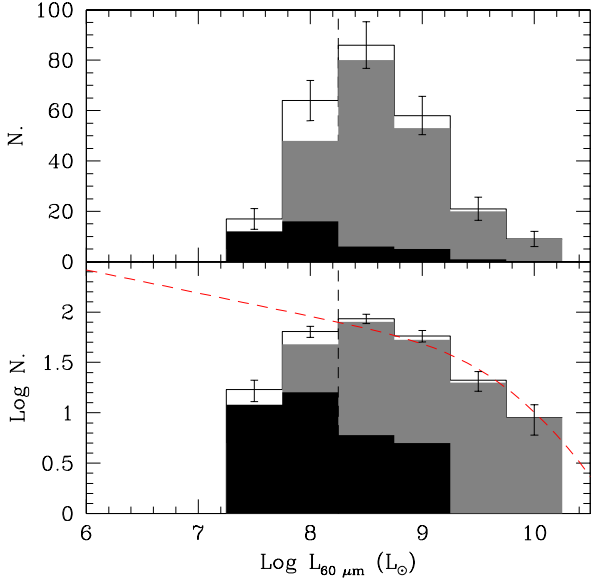}
\small{\caption{The infrared luminosity distribution in linear (upper panel) and logarithmic (lower panel) scales
for the FIR (60 $\mu$m) detected HRS galaxies (empty histogram).
The black and gray histograms are for early-type (E-S0-S0a) and late-type (Sa-Sd-Im-BCD) galaxies respectively.
The dashed line is the IRAS 60 $\mu$m luminosity function determined by Takeuchi et al. (2003).
The vertical, dashed line indicates the typical detection limit of IRAS
(0.4 mJy) at the distance of 20 Mpc.
\label{statfir}}}
\end{figure}

Figure \ref{statfir} shows the far-IR luminosity distribution of the HRS early-type (black histogram) and
late-type (gray histogram) galaxies detected by IRAS at 60  $\mu$m compared the 
IRAS 60 $\mu$m luminosity function determined by Takeuchi et al. (2003) ($\alpha$ = 1.23 $\pm$ 0.04, $L^*$ = 8.85 $\times$ 10$^8$ L$\odot$
and $\phi$ $^*$ in arbitrary units). 
The luminosity distribution of the HRS follows the luminosity function over two orders 
of magnitude. The low-luminosity cutoff is due to the detection limit of the IRAS survey. 
The lack of very luminous galaxies is because Luminous (LIRGs, $L_{60 \mu m}$ = 10 $^{11}$ L$\odot$) 
and Ultra Luminous (ULIRGs, $L_{60 \mu m}$ = 10 $^{12}$ L$\odot$) Infrared Galaxies 
are quite rare in the nearby universe and in particular inside rich clusters (Boselli \& Gavazzi 2006). \\

\begin{figure}
\epsscale{1.5} 
\plotone{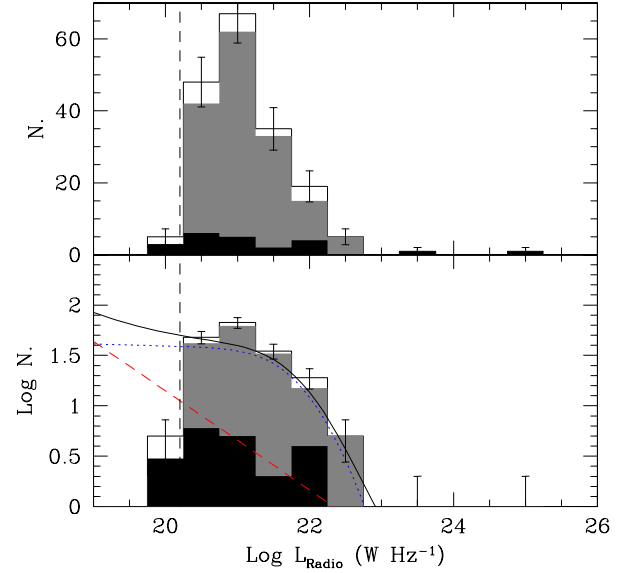}
\small{\caption{The radio continuum (1415 MHz) luminosity distribution in linear (upper panel) and logarithmic (lower panel) scales
for the detected HRS galaxies (empty histogram).
The black and gray histograms are for early-type (E-S0-S0a) and late-type (Sa-Sd-Im-BCD) galaxies respectively
compared to the 2dF/NVSS luminosity function of Mauch \& Sadler (2007) (solid line). This radio luminosity function
includes both the contribution of AGN (red dashed line) and quiescent (blue dotted line) galaxies.
The vertical, dashed line indicates the typical detection limit of NVSS
(2.5 mJy) at the distance of 20 Mpc.
\label{statrad}}}
\end{figure}

Figure \ref{statrad} shows the radio continuum (1.4 GHz) luminosity distribution of the radio 
detected HRS early-type (black histogram) and late-type (gray histogram) galaxies, as
compared to the 2dF/NVSS luminosity function of Mauch \& Sadler (2007) (solid line). This radio luminosity function
includes the contribution from both AGN (dashed line) and quiescent (dotted line) galaxies. The radio luminosity distribution of the
HRS is quite typical of normal, nearby galaxies (10$^{20}$ $\leq$ $L_{Radio}$ $\leq$ 10$^{22}$ W Hz$^{-1}$; Condon et al. 2002). The two brightest 
sources are the radio galaxies M87 (Virgo A, $L_{Radio}$ = 10$^{24.8}$ W Hz$^{-1}$) and M84 ($L_{Radio}$ = 10$^{23.3}$ W Hz$^{-1}$).
The radio continuum luminosity distribution as a whole agrees well with the field luminosity function, indicating that the contamination of
cluster galaxies with an enhanced radio continuum emission is minimal (Gavazzi \& Boselli 1999c,d).

In summary, since the K-band light traces stellar mass,
the good agreement between the HRS luminosity distribution and K-band luminosity function implies that
the HRS can be treated as a volume-limited sample down to a minimum stellar mass. Figure \ref{statfir} shows that
the HRS is also representative of a far-IR selected sample once the most active and rare
in the local universe LIRG and ULIRG are excluded. The HRS matches also the distribution of a radio continuum
selected sample, although the statistic for the brightest radio galaxies is poor.
However, Fig. \ref{statHI} shows that when viewed from the perspective of gas mass, 
the HRS is not a fair sample of the local universe. The presence of HI-deficient objects makes it ideal for
studying the effects of the cluster environment on the dust properties of nearby galaxies.

\acknowledgements
We would like to thank the Herschel Project Scientist, G. Pilbratt, the SPIRE team and all the people involved in the
construction and the launch of Herschel.
This research has made use of the NASA/IPAC Extragalactic Database (NED) which is operated by the 
Jet Propulsion Laboratory, California Institute of Technology, under contract with the National 
Aeronautics and Space Administration, and of the GOLD Mine database. A.B. wishes to thank S. Boissier 
for his help in the construction of the HRS database. We are grateful to the anonymous referee 
for his precious comments and suggestions which helped improving the quality of the manuscript.

\newpage

\references

\reference{}Adelman-McCarthy J., et al., 2008, ApJS, 175, 297

\reference{}Alton, P., Trewhella, M., Davies, J., et al., 1998, A\&A, 335, 807

\reference{}Alton, P., Trewhella, M., Davies, J., Bianchi, S., 1999, A\&A, 343, 51

\reference{}Beker, R., White, R., Helfand, D., 1995, ApJ, 450, 559

\reference{}Bendo, G., Joseph, R., Wells, M., et al., 2003, AJ, 125, 2361

\reference{}Binggeli B., Sandage A., Tammann G., 1985, AJ, 90, 1681

\reference{}B\"ohringer, H., Briel, U. G., Schwarz, R. A., Voges, W., Hartner, G., Trumper, J., 1994, Nature, 368, 828
    
\reference{}Boselli, A. \& Gavazzi, G., 2002, A\&A, 386, 124

\reference{}Boselli, A. \& Gavazzi, G., 2006, PASP, 118, 517

\reference{}Boselli, A. \& Gavazzi, G., 2009, A\&A, 508, 201

\reference{}Boselli A., Casoli F., Lequeux J., 1995, A\&AS, 110, 521

\reference{}Boselli A., Tuffs R., Gavazzi G., Hippelein H., Pierini D., 1997, A\&A 121, 507

\reference{}Boselli, A., Gavazzi, G., Franzetti, P., Pierini, D., Scodeggio, M., 2000, A\&AS, 142, 73

\reference{}Boselli, A., Gavazzi, G., Donas, J., Scodeggio, M., 2001, AJ, 121, 753

\reference{}Boselli A., Lequeux J., Gavazzi G., 2002, A\&A, 384, 33

\reference{}Boselli, A., Iglesias-P\'aramo, J., Vilchez, J.M., Gavazzi, G., 2002, A\&A, 386, 134

\reference{}Boselli A., Gavazzi G., Sanvito G., 2003, A\&A, 402, 37

\reference{}Boselli A., Sauvage M., Lequeux J., Donati A., Gavazzi G., 2003, A\&A, 406, 867

\reference{}Boselli A., Lequeux J., Gavazzi G., 2004, A\&A, 428, 409

\reference{}Boselli, A., Cortese, L., Deharveng, JM., et al., 2005a, ApJ, 629, L29

\reference{}Bower, R.G., Lucey, J.R., Ellis, R.S., 1992, MNRAS, 254, 601

\reference{}Bregman, J., Snider, B., Grego, L., Cox, C., 1998, ApJ, 499, 670

\reference{}Bressan, A., Granato, GL., Silva, L., 1998, A\&A, 332, 135

\reference{}Bressan, A., Panuzzo, P., Buson, L., et al., 2006, ApJ, 639, L55

\reference{}Broelis, A.H., \& Van Woerden, H., 1994, A\&AS, 107, 129

\reference{}Buat, V., \& Xu, K., 1996, A\&A, 306, 61

\reference{}Buat, V., Boselli, A., Gavazzi, G., Bonfanti, C., 2002, A\&A, 383, 801

\reference{}Calzetti, D., Kennicutt, R., Bianchi, L., et al., 2005, ApJ, 633, 871

\reference{}Calzetti, D., Kennicutt, R., Engelbracht, C., et al., 2007, ApJ, 666, 870

\reference{}Catinella, B., Giovanelli, R., Haynes, M., 2006, ApJ, 640, 751

\reference{}Cayatte, V., Kotanyi, C., Balkowski, C., \& van Gorkom, J., 1994, AJ, 107, 1003

\reference{}Cayatte, V., van Gorkom, J., Balkowski, C., \& Kotanyi, C., 1990, AJ, 100, 604

\reference{}Chanial, P., Flores, H., Guiderdoni, B., Elbaz, D., Hammer, F., Vigroux, L., 2007, A\&A, 462, 81

\reference{}Chung, A., van Gorkom, J., Kenney, J., Crowl, H., Vollmer, B., 2009, AJ, 138, 1741

\reference{}Cole, S., Norberg, P., Baugh,, C., et al., 2001, MNRAS, 326, 255 

\reference{}Condon, J. J., Cotton, W. D., Broderick, J. J., 2002, AJ, 124, 675

\reference{}Condon, J. J., Cotton, W. D., Greisen, E. W., Yin, Q. F., Perley, R. A., Taylor, G. B., Broderick, J. J. 1998, AJ, 115, 1693

\reference{}Cortese, L., Boselli, A., Buat, V., et al., 2006, ApJ, 637, 242

\reference{}Cortese, L., Boselli, A., Franzetti, P., Decarli, R., Gavazzi, G., Boissier, S., Buat, V., 2008a, MNRAS, 386, 1157

\reference{}Cortese, L., Minchin, R., Auld, R., et al., 2008b, MNRAS, 383, 1519

\reference{}Cortese, L., Hughes, T., 2009, MNRAS, 400, 1225

\reference{}Cuillandre, J.C., Lequeux, J., Allen, R., Mellier, Y., Bertin, E., 2001, ApJ, 554, 190

\reference{}Dale, D., Bendo, G., Engelbracht, C., et al., 2005, ApJ, 633, 857

\reference{}Dale, D., Gil de Paz, A., Gordon, K., et al., 2007, ApJ, 655, 863

\reference{}Davies, J., Alton, P., Bianchi, S., Trewhella, M., 1998, MNRAS, 300, 1006

\reference{}Davies, J., Alton, P., Trewhella, M., Evans, R., Bianchi, S., 1999, MNRAS, 304, 495

\reference{}D\'esert F.--X., Boulanger F., Puget J.--L. 1990, A\&A 237, 215

\reference{}De Vaucouleurs, G., De Vaucouleurs, A., Corwin, H., Buta, R., Paturel, G., Fouque, P., 1991, Third Reference Catalogue of Bright Galaxies

\reference{}Devereux, N., \& Young, J., 1990, ApJ, 359, 42

\reference{}Djorgovski, S., Davis, M., 1987, ApJ, 313, 59

\reference{}Draine, B., 1978, ApJS, 36, 596

\reference{}Draine, B., \& Li, A., 2007, ApJ, 657, 810

\reference{}Draine, B., Dale, D., Bendo, G., et al., 2007, ApJ, 663, 866

\reference{}Dressler, A., Lynden-Bell, D., Burstein, D., Davies, R., Faber, S., Terlevich, R., Wegner, G., 1987, ApJ, 313, 42

\reference{}Duley, W., Williams, D., 1986, MNRAS, 223, 177

\reference{}Dunne, L., Eales, S., 2001, MNRAS, 327, 697

\reference{}Dunne, L., Eales, S., Edmunds, M., Ivison, R., Alexander, P., Clements, D., 2000, MNRAS, 315, 115

\reference{}Dwek E., 1986, ApJ, 302, 363

\reference{}Dwek E., Arendt R., Fixsen D., et al., 1997, ApJ, 475, 565

\reference{}Eales, S., Wynn-Williams, C., Duncan, W., 1989, ApJ, 339, 859

\reference{}Engelbracht, C., Kundurthy, P., Gordon, K., et al., 2006, ApJ, 642, L127

\reference{}Fabbiano, G., Kim, D., Trinchieri, G., 1992, ApJS,80, 531

\reference{}Ferrarese, L., C\^ot\'e, P., Jord\'an, A., et al., 2006, ApJS, 164, 334

\reference{}Ferrari, F., Pastoriza, M.G., Macchetto, F., Caon, N., 1999, A\&AS, 136, 269

\reference{}Galliano, F., Madden, S., Jones, A., Wilson, C., Bernard, J-P., Le Peintre, F., 2003, A\&A, 407, 159

\reference{}Galliano, F., Madden, S., Jones, A., Wilson, C., Bernard, J-P., 2005, A\&A, 434, 867

\reference{}Galliano, F., Dwek, E., Chanial, P. 2008, ApJ, 672, 214

\reference{}Gallo, E., Treu, T., Jacob, J., Woo, JH, Marshall, P., Antonucci, R., 2008, ApJ, 680, 154

\reference{}Garcia-Barreto, J., Dowens, D., Huchtmeier, W., 1994, A\&A, 288, 705

\reference{}Gavazzi, G., Pierini, D. \& Boselli, A., 1996, A\&A, 312, 397

\reference{}Gavazzi, G., Catinella, B., Carrasco, L., Boselli, A., Contursi, A., 1998, AJ, 115, 1745

\reference{}Gavazzi, G., Boselli, A., Scodeggio, M., Pierini, D., Belsole, E., 1999a, MNRAS, 304, 595

\reference{}Gavazzi, G.,Carrasco, L., Galli, R., 1999b, A\&AS, 136, 227

\reference{}Gavazzi, G., Boselli, A., 1999c, A\&A, 343, 86

\reference{}Gavazzi, G., Boselli, A., 1999d, A\&A, 343, 93

\reference{}Gavazzi, G., Franzetti, P., Scodeggio, M., Boselli, A., Pierini, D., 2000, A\&A, 361, 863

\reference{}Gavazzi, G., Franzetti, P., Scodeggio, M., Boselli, A., Pierini, D., Baffa, C., Lisi, F., Hunt, L., 2000, A\&AS, 142, 65

\reference{}Gavazzi, G., Boselli, A., Pedotti, P., Gallazzi, A., Carrasco, L., 2002, A\&A, 396, 449

\reference{}Gavazzi, G., Bonfanti, C., Sanvito, G., Boselli, A., \& Scodeggio, M., 2002b, ApJ, 576, 135

\reference{}Gavazzi, G., Boselli, A., Donati, A., Franzetti, P. \& Scodeggio, M., 2003, A\&A, 400, 451

\reference{}Gavazzi, G., Zaccardo, A., Sanvito, G., Boselli, A. \& Bonfanti, C., 2004, A\&A, 417, 499

\reference{}Gavazzi, G., Boselli, A., van Driel, W., O'Neil, K., 2005, A\&A, 429, 439

\reference{}Gavazzi, G., Boselli, A., Cortese, L., Arosio, I, Gallazzi, A., Pedotti, P., 2006, A\&A, 446, 839

\reference{}Gavazzi, G., Giovanelli, R., Haynes, M., et al., 2008, A\&A, 482, 43

\reference{}Gerin, M., Casoli, F., 1994, A\&A, 290, 49

\reference{}Giovanelli, R., Haynes, M., Kent, B., et al., 2005, AJ, 130, 25, 98

\reference{}Gil de Paz, A., Boissier, S., Madore, B., et al., 2007, ApJS, 173, 185

\reference{}Gordon, K., Perez-Gonzalez, P., Misselt, K., et al., 2004, ApJS, 154, 215

\reference{}Gordon, K., Rieke, G. H., Engelbracht, C. W., et al., 2005, PASP, 117, 831

\reference{}Goudfrooij, P., Hansen, L., Jorgensen, H.E., Norgaard-Nielsen, H.U., 1994, A\&AS, 105, 341 

\reference{}Griffin, M., Abergel, A., Ade, P., Andr\'e, P., Baluteau, J-P., Bock, J., Franceschini, A., Gear, W., et al., 2007, Advances in Space Research, Vol 40, Issue 5, p.612-619

\reference{}Griffin, M., Abergel, A., Ade, P., Andr\'e, P., Baluteau, J-P., Bock, J., Franceschini, A., Gear, W., et al., 2006, SPIE, 6265, 7

\reference{}Gu\'elin M., Zylka R., Mezger P.G. et al., 1993, A\&A 279, L37

\reference{}Gu\'elin M., Zylka R., Mezger P.G., Haslam C.G.T., Kreysa, E., 1995, A\&A, 298, L29

\reference{}James, A., Dunne, L., Eales, S., Edmunds, M., 2002, MNRAS, 335, 753

\reference{}Jansen, R., Fabricant, D., Franx, M., Caldwell, N., 2000, ApJS, 126, 331

\reference{}Jarrett, T., Chester, T., Cutri, R., Schneider, S. \& Huchra, J., 2003, AJ, 125, 525

\reference{}Haynes, M., \& Giovanelli, R., 1984, AJ, 89, 758

\reference{}Heavens, A., Panter, B., Jimenez, R., Dunlop, J., 2004, Nat, 428, 625

\reference{}Hirashita, H., Buat, V., Inoue, A., 2003, A\&A, 410, 83

\reference{}Hollenbach, D., Salpeter, E., 1971, ApJ, 163, 155

\reference{}Hollenbach, D., Tielens, A., 1997, ARA\&A, 35, 179

\reference{}Holmberg, E., 1958, Lund Medd. Astron. Obs. Ser. II, 136, 1

\reference{}Hughes, T., Cortese, L., 2009, MNRAS, 396, L41

\reference{}Kaneda, H., Onaka, T., Kitayama, T., Okada, Y., Sakon, I., 2007, PASJ, 59, 107

\reference{}Karachentsev, I., Lebedev, V., Shcherbanovskij, A., 1972, Publ. Special Astrophys. Obs. of USSR 7

\reference{}Kauffmann, G., Heckman, T., White, S., et al., 2003, MNRAS, 341, 54

\reference{}Kenney J.,Young J., 1988, ApJS 66, 261

\reference{}Kennicutt, R., 1992a, ApJ, 388, 310

\reference{}Kennicutt, R., 1992b, ApJS, 79, 255

\reference{}Kennicutt, R., 1998, ARA\&A, 36, 189

\reference{}Kennicutt, R., Armus, L., Bendo, G., et al., 2003, PASP, 115, 928

\reference{}Kennicutt, R., Calzetti, D., Walter, F., et al., 2007, ApJ, 671, 333

\reference{}Kochanek, C., Pahre, M., Falco, E., Huchra, J., Mader, J., Jarrett, T., Chester, T., Cutri, R., Schneider, S., 2001, ApJ, 560, 566

\reference{}Leech K, V\"olk H., Heinrichsen I., et al., 1999, MNRAS, 310, 317

\reference{}Leeuw, L., Hawarden, T., Matthews, H., Robson, I., Eckart, A., 2002, ApJ, 565, L131

\reference{}Leeuw, L., Davidson, J., Dowell, K., Matthews, H., 2008, ApJ, 677, L249

\reference{}Malhotra, S., Kaufman, M., Hollenbach, D., et al., 2001, ApJ, 561, 766

\reference{}Mauch, T, \& Sadler, E., 2007, MNRAS, 375, 931

\reference{}Meiksin, A., 2009, Rev. Mod. Phys., 81, 1405

\reference{}Montier, L., Giard, M., 2005, A\&A, 439, 35

\reference{}Morton, R., \& Haynes, M., 1994, ARA\&A, 32, 115

\reference{}Moustakas, J., \& Kennicutt, R., 2006, ApJS, 164, 81

\reference{}Neininger N., Gu\'elin M., Garcia-Burillo S., Zylka R., Wielebinski R., 1996, A\&A 310, 725

\reference{}Niklas, S., Klein, U., Wielebinski, R., 1995, A\&A, 293, 56

\reference{}Nilson, P., 1973, The Uppsala General Catalogue of Galaxies, 1973, Uppsala University Press 

\reference{}Nolthenius, R., 1993, ApJS, 85, 1

\reference{}Oppenheimer, B., Dav\'e, R., 2008, MNRAS, 387, 577

\reference{}Panuzzo, P., Vega, O., Bressan, A., et al., 2007, ApJ, 656, 206

\reference{}Perez-Gonzalez, P., Kennicutt, R., Gordon, K., et al., 2006, ApJ, 648, 987

\reference{}Prescott, M., Kennicutt, R., Bendo, G., et al., 2007, ApJ, 668, 182

\reference{}Rayan-Weber, E., Pettini, M., Madau, P., 2006, MNRAS, 371, L78

\reference{}Roussel H., Vigroux L., Bosma A., et al., 2001b, A\&A, 369, 473

\reference{}Sarazin, C., 1986, Rev Mod Phys, 58, 1

\reference{}Schlegel, F., Finkbeiner, D., Davis M., 1998, ApJ, 500, 525

\reference{}Scodeggio, M., Gavazzi, G., Franzetti, P., Boselli, A., Zibetti, S., Pierini, D., 2002, A\&A, 384, 812

\reference{}Shapley, A., Fabbiano, G., Eskridge, P., 2001, ApJS, 137, 139

\reference{}Skrutskie, M., Cutri, R., Stiening, R., et al., 2006, AJ, 131, 1163

\reference{}Soifer, B., Neugebauer, G., Houck, J., 1987, ARA\&A, 25, 187

\reference{}Springob, C., Haynes, M., Giovanelli, R., Kent, B., 2005, ApJS, 160, 149

\reference{}Stickel, M., Klaas, U., Lemke, D., Mattila, D., 2002, A\&A, 383, 367

\reference{}Takeuchi, T., Yoshikawa, K., Ishii, T., 2003, ApJ, 587, L89

\reference{}Temi, P., Brighenti, F., Mathews, W., 2007, ApJ, 660, 1215

\reference{}Tielens, A. McKee, C., Seab, C., Hollenbach, D., 1994, ApJ, 431, 321

\reference{}Thilker, D., Boissier, S., Bianchi, L., et al., 2007, ApJS, 173, 572

\reference{}Thomas, H., Dunne, L., Clemens, M., Alexander, P., Eales, S., Green, D., James, A., 2002, MNRAS, 331, 853

\reference{}Tremonti, C., Heckman, T., Kauffmann, G., et al., 2004, ApJ, 613, 898

\reference{}Tuffs R., Popescu C., Pierini D., et al., 2002, ApJS, 139, 37

\reference{}Tully, B., Fisher, J., 1977, A\&A, 54, 661

\reference{}Tully, B., Mould, J., Aaronson, M., 1982, ApJ, 257, 527

\reference{}Tully, B., 1988, "Nearby Galaxy Catalog", Cambridge University Press

\reference{}van der Hulst, J., 2002, in Seeing Through the Dust: The Detection of HI and the Exploration of the ISM in Galaxies, ASP Conference Proceedings, Vol. 276. Edited by A. R. Taylor, T. L. Landecker, and A. G. Willis. ISBN: 1-58381-118-4. San Francisco: Astronomical Society of the Pacific, 2002., p.84

\reference{}Van Dokkum, P.G., \& Franx, M., 1995, ApJ, 110, 2027

\reference{}Verheijen, M., Sancisi, R., 2001, A\&A, 370, 765

\reference{}Visvanathan, N., Sandage, A., 1977, ApJ, 216, 214

\reference{}Vlahakis, C., Dunne, L., Eales, S., 2005, MNRAS, 364, 1253

\reference{}Vollmer, B., Thierbach, M., Wielebinski, R., 2004, A\&A, 418, 1

\reference{}Ward-Thompson, D., Andr\'e, P., Kirk, J., 2002, MNRAS, 329, 257


\reference{}Waskett, T., Sibthorpe, B., Griffin J., Chanial, P.F., 2007, MNRAS, 381, 1583

\reference{}Whittet, D., in "Dust in the galactic environment", The Graduate Series in Astronomy, Bristol: Institute of Physics (IOP) Publishing, 1992

\reference{}Witt, A., \& Gordon, K., 2000, ApJ, 528, 799

\reference{}Xilouris, E., Madden, S., Galliano, F., Virgoux, L., Sauvage, M., 2004, A\&A, 416, 41

\reference{}Young, J., Xie, S., Tacconi, L., et al., 1995, ApJS 98, 219

\reference{}Young, L., Bendo, G., Lucero, D., 2008, in "The Evolving ISM in the Milky Way and Nearby Galaxies," the 4th Spitzer Science Center Conference (Pasadena, December 2007

\reference{}Zaritsky, D., Kennicutt, R., Huchra, J., 1994, ApJ, 420, 87

\reference{}Zubko, V., Dwek, E., Arendt, R., 2004, ApJS, 152, 211

\reference{}Zwaan, M., Meyer, M., Staveley-Smith, L., Webster, R., 2005, MNRAS, 359, L30

\reference{}Zwicky F., Herzog E., Karpowicz M., Kowal C., Wild P., 1961-1968, "Catalogue of Galaxies and of Cluster of Galaxies" (Pasadena, California Institute of Technology; CGCG)

\clearpage

\begin{table*}
\tiny
\caption{Ancillary data for the Herschel Reference Sample.}
\label{Tabdata}
\[
\begin{tabular}{rccccccccc}
\hline
\noalign{\smallskip}
     HRS& 2MASS & SDSS & UV & H$\alpha$ & IRAS & 20cm & HI & CO & Spec \\
\noalign{\smallskip}    
\hline
\noalign{\smallskip}          
       1   & X &   X  &	- &    -  &  X  &    -  &    -  &    -  &    X  \\  
       2   & X &   X  &	X &    -  &  X  &    X  &    X  &    -  &    X  \\  
       3   & X &   X  &	- &    X  &  -  &    X  &    X  &    X  &    -  \\  
       4   & X &   X  &	- &    X  &  X  &    X  &    X  &    X  &    X  \\  
       5   & X &   X  &	- &    X  &  X  &    X  &    X  &    -  &    X  \\  
       6   & X &   X  &	X &    -  &  -  &    -  &    X  &    -  &    -  \\  
       7   & X &   X  &	X &    -  &  X  &    X  &    X  &    X  &    -  \\  
       8   & X &   X  &	X &    X  &  X  &    -  &    X  &    X  &    X  \\  
       9   & X &   X  &	X &    X  &  X  &    X  &    X  &    X  &    X  \\  
      10   & X &   X  &	- &    -  &  X  &    X  &    X  &    -  &    X  \\  
      11   & X &   X  &	X &    X  &  X  &    X  &    X  &    X  &    X  \\  
      12   & X &   X  &	X &    -  &  X  &    -  &    X  &    -  &    X  \\  
      13   & X &   X  &	X &    X  &  X  &    X  &    X  &    X  &    X  \\  
      14   & X &   X  &	X &    -  &  X  &    X  &    X  &    -  &    -  \\  
      15   & X &   X  &	X &    X  &  X  &    X  &    X  &    X  &    X  \\  
      16   & X &   X  &	X &    X  &  X  &    -  &    X  &    X  &    X  \\  
      17   & X &   X  &	X &    X  &  X  &    X  &    X  &    X  &    X  \\  
      18   & X &   X  &	X &    X  &  X  &    X  &    X  &    -  &    X  \\  
      19   & X &   X  &	X &    X  &  X  &    X  &    X  &    -  &    X  \\  
      20   & X &   X  &	X &    X  &  -  &    X  &    X  &    X  &    X  \\  
      21   & X &   X  &	X &    X  &  -  &    -  &    X  &    -  &    X  \\  
      22   & X &   X  &	X &    X  &  -  &    X  &    X  &    -  &    -  \\  
      23   & X &   X  &	X &    X  &  X  &    X  &    X  &    X  &    X  \\  
      24   & X &   X  &	X &    X  &  X  &    X  &    X  &    X  &    X  \\  
      25   & X &   X  &	X &    X  &  X  &    X  &    X  &    X  &    X  \\  
      26   & X &   X  &	X &    -  &  X  &    -  &    X  &    -  &    X  \\  
      27   & X &   X  &	X &    X  &  X  &    X  &    X  &    -  &    X  \\  
      28   & X &   X  &	X &    X  &  X  &    X  &    X  &    -  &    X  \\  
      29   & X &   X  &	X &    -  &  -  &    -  &    X  &    -  &    X  \\  
      30   & X &   X  &	X &    -  &  -  &    X  &    X  &    -  &    X  \\  
      31   & X &   X  &	X &    X  &  X  &    X  &    X  &    X  &    X  \\  
      32   & X &   X  &	X &    X  &  -  &    -  &    X  &    X  &    X  \\  
      33   & X &   X  &	X &    X  &  -  &    X  &    X  &    X  &    X  \\  
      34   & X &   X  &	X &    X  &  -  &    X  &    X  &    X  &    X  \\  
      35   & X &   X  &	- &    -  &  -  &    -  &    -  &    -  &    X  \\  
      36   & X &   X  &	X &    X  &  X  &    X  &    X  &    X  &    X  \\  
      37   & X &   X  &	X &    X  &  X  &    X  &    X  &    X  &    X  \\  
      38   & X &   X  &	- &    X  &  X  &    X  &    X  &    X  &    X  \\  
      39   & X &   X  &	X &    -  &  -  &    -  &    X  &    -  &    X  \\  
      40   & X &   X  &	X &    X  &  -  &    X  &    X  &    -  &    X  \\  
      41   & X &   X  &	X &    X  &  -  &    -  &    X  &    -  &    -  \\  
      42   & X &   X  &	- &    X  &  -  &    X  &    X  &    X  &    X  \\  
      43   & X &   X  &	X &    -  &  -  &    -  &    X  &    -  &    -  \\  
      44   & X &   X  &	X &    -  &  -  &    X  &    X  &    -  &    X  \\  
      45   & X &   X  &	X &    -  &  X  &    X  &    X  &    -  &    -  \\  
      46   & X &   X  &	- &    -  &  -  &    X  &    X  &    -  &    -  \\  
      47   & X &   X  &	X &    X  &  -  &    X  &    X  &    -  &    X  \\  
      48   & X &   X  &	X &    X  &  X  &    X  &    X  &    X  &    X  \\  
      49   & X &   X  &	X &    -  &  -  &    -  &    X  &    -  &    -  \\  
      50   & X &   X  &	X &    X  &  X  &    X  &    X  &    X  &    X  \\  
      51   & X &   X  &	- &    X  &  X  &    X  &    X  &    -  &    X  \\  
      52   & X &   X  &	X &    -  &  X  &    X  &    X  &    -  &    X  \\  
      53   & X &   X  &	X &    X  &  X  &    X  &    X  &    X  &    X  \\  
      54   & X &   X  &	X &    X  &  X  &    -  &    X  &    X  &    X  \\  
      55   & X &   X  &	X &    X  &  X  &    X  &    X  &    X  &    X  \\  
      56   & X &   X  &	X &    X  &  X  &    X  &    X  &    X  &    X  \\  
      57   & X &   X  &	X &    X  &  X  &    X  &    X  &    X  &    X  \\  
      58   & X &   X  &	X &    X  &  X  &    X  &    X  &    -  &    X  \\  
      59   & X &   X  &	X &    X  &  X  &    X  &    X  &    X  &    X  \\  
      60   & X &   X  &	X &    X  &  X  &    X  &    X  &    X  &    X  \\  
      61   & X &   X  &	X &    -  &  X  &    -  &    X  &    -  &    X  \\  
      62   & X &   X  &	X &    X  &  -  &    X  &    X  &    -  &    X  \\  
      63   & X &   X  &	- &    X  &  X  &    -  &    X  &    X  &    X  \\  
      64   & X &   X  &	- &    -  &  X  &    -  &    X  &    -  &    X  \\  
      65   & X &   X  &	X &    -  &  X  &    -  &    X  &    -  &    X  \\  
      66   & X &   X  &	X &    X  &  -  &    X  &    X  &    X  &    X  \\  
      67   & X &   X  &	- &    -  &  -  &    X  &    X  &    -  &    X  \\  
      68   & X &   X  &	X &    -  &  -  &    X  &    X  &    -  &    X  \\  
      69   & X &   X  &	- &    X  &  X  &    X  &    X  &    X  &    X  \\  
      70   & X &   -  &	X &    X  &  X  &    X  &    X  &    -  &    X  \\  
      71   & X &   X  &	X &    -  &  X  &    -  &    -  &    -  &    -  \\  
      72   & X &   -  &	X &    -  &  X  &    X  &    X  &    -  &    X  \\  
      73   & X &   X  &	- &    X  &  X  &    -  &    X  &    X  &    X  \\  
      74   & X &   X  &	- &    X  &  X  &    X  &    X  &    X  &    X  \\  
      75   & X &   X  &	- &    -  &  -  &    -  &    X  &    -  &    X  \\  
      76   & X &   X  &	- &    X  &  X  &    -  &    X  &    -  &    X  \\  
      77   & X &   X  &	X &    X  &  X  &    X  &    X  &    X  &    X  \\  
      78   & X &   X  &	X &    X  &  X  &    -  &    X  &    X  &    X  \\  
      79   & X &   X  &	X &    X  &  X  &    X  &    X  &    -  &    X  \\  
      80   & X &   X  &	X &    X  &  X  &    -  &    X  &    -  &    X  \\  
      81   & X &   X  &	X &    X  &  X  &    X  &    X  &    X  &    X  \\  
      82   & X &   X  &	X &    -  &  X  &    X  &    X  &    -  &    X  \\  
      83   & X &   X  &	X &    -  &  X  &    -  &    X  &    -  &    X  \\  
      84   & X &   X  &	X &    X  &  X  &    X  &    X  &    X  &    X  \\  
      85   & X &   X  &	X &    X  &  X  &    X  &    X  &    X  &    X  \\  
      86   & X &   X  &	X &    X  &  X  &    X  &    X  &    X  &    X  \\  
      87   & X &   X  &	X &    -  &  X  &    -  &    X  &    -  &    X  \\  
      88   & X &   X  &	X &    X  &  X  &    X  &    X  &    X  &    X  \\  
      89   & X &   X  &	X &    X  &  X  &    X  &    X  &    X  &    X  \\  
      90   & X &   X  &	X &    -  &  -  &    -  &    X  &    -  &    X  \\  
\noalign{\smallskip}
\hline
\end{tabular}
\]
\end{table*}

\addtocounter{table}{-1}
\begin{table*}
\tiny
\caption{continue}
\label{Tabmod}
\[\begin{tabular}{rccccccccc}
\hline
\noalign{\smallskip}
     HRS& 2MASS & SDSS & UV & H$\alpha$ & IRAS & 20cm & HI & CO & Spec \\
\noalign{\smallskip}    
\hline
\noalign{\smallskip}          
      91   & X &   X  &	- &    X  &  X  &    X  &    X  &    X  &    X  \\  
      92   & X &   X  &	- &    X  &  X  &    -  &    X  &    -  &    X  \\  
      93   & X &   X  &	X &    -  &  X  &    X  &    X  &    X  &    -  \\  
      94   & X &   X  &	X &    X  &  X  &    -  &    X  &    X  &    X  \\  
      95   & X &   X  &	X &    X  &  X  &    X  &    X  &    X  &    X  \\  
      96   & X &   X  &	- &    X  &  X  &    X  &    X  &    X  &    X  \\  
      97   & X &   X  &	X &    X  &  X  &    X  &    X  &    X  &    X  \\  
      98   & X &   X  &	X &    X  &  X  &    X  &    X  &    X  &    X  \\  
      99   & X &   X  &	X &    X  &  X  &    -  &    X  &    -  &    X  \\  
     100   & X &   X  &	X &    X  &  X  &    X  &    X  &    X  &    X  \\  
     101   & X &   X  &	X &    -  &  X  &    -  &    X  &    X  &    -  \\  
     102   & X &   X  &	X &    X  &  X  &    X  &    X  &    X  &    X  \\  
     103   & X &   X  &	X &    X  &  X  &    -  &    X  &    X  &    X  \\  
     104   & X &   X  &	X &    X  &  -  &    -  &    -  &    -  &    -  \\  
     105   & X &   X  &	- &    -  &  -  &    -  &    X  &    -  &    X  \\  
     106   & X &   X  &	X &    X  &  X  &    -  &    X  &    -  &    X  \\  
     107   & X &   X  &	X &    X  &  -  &    -  &    X  &    -  &    X  \\  
     108   & X &   X  &	X &    X  &  -  &    -  &    X  &    -  &    X  \\  
     109   & X &   X  &	X &    X  &  X  &    X  &    X  &    -  &    X  \\  
     110   & X &   X  &	X &    X  &  X  &    X  &    X  &    X  &    X  \\  
     111   & X &   X  &	X &    X  &  X  &    X  &    X  &    X  &    X  \\  
     112   & X &   X  &	- &    X  &  X  &    -  &    X  &    X  &    X  \\  
     113   & X &   X  &	X &    X  &  X  &    X  &    X  &    X  &    X  \\  
     114   & X &   X  &	X &    X  &  X  &    X  &    X  &    X  &    X  \\  
     115   & X &   X  &	X &    X  &  -  &    -  &    X  &    -  &    X  \\  
     116   & X &   X  &	- &    X  &  -  &    -  &    X  &    -  &    X  \\  
     117   & X &   X  &	- &    X  &  X  &    X  &    X  &    X  &    X  \\  
     118   & X &   X  &	X &    X  &  X  &    -  &    X  &    -  &    X  \\  
     119   & X &   X  &	X &    X  &  X  &    X  &    X  &    X  &    X  \\  
     120   & X &   X  &	X &    X  &  X  &    -  &    X  &    X  &    X  \\  
     121   & X &   X  &	X &    X  &  X  &    X  &    X  &    X  &    X  \\  
     122   & X &   X  &	X &    X  &  X  &    X  &    X  &    X  &    X  \\  
     123   & X &   X  &	X &    X  &  X  &    -  &    X  &    -  &    X  \\  
     124   & X &   X  &	X &    X  &  X  &    X  &    X  &    X  &    X  \\  
     125   & X &   X  &	X &    -  &  -  &    -  &    X  &    -  &    -  \\  
     126   & X &   X  &	- &    -  &  X  &    -  &    X  &    -  &    -  \\  
     127   & X &   X  &	X &    X  &  X  &    X  &    X  &    X  &    X  \\  
     128   & X &   X  &	X &    X  &  X  &    -  &    X  &    -  &    X  \\  
     129   & X &   X  &	X &    -  &  X  &    -  &    X  &    -  &    X  \\  
     130   & X &   X  &	X &    X  &  X  &    -  &    X  &    X  &    X  \\  
     131   & X &   X  &	X &    X  &  -  &    -  &    X  &    -  &    X  \\  
     132   & X &   X  &	X &    X  &  X  &    X  &    X  &    -  &    X  \\  
     133   & X &   X  &	X &    X  &  X  &    -  &    X  &    X  &    X  \\  
     134   & X &   X  &	- &    X  &  X  &    -  &    X  &    -  &    X  \\  
     135   & X &   X  &	X &    X  &  -  &    -  &    -  &    -  &    X  \\  
     136   & X &   X  &	- &    X  &  X  &    -  &    X  &    X  &    X  \\  
     137   & X &   X  &	X &    -  &  -  &    -  &    X  &    -  &    X  \\  
     138   & X &   X  &	X &    X  &  X  &    X  &    X  &    X  &    -  \\  
     139   & X &   X  &	X &    X  &  X  &    -  &    X  &    -  &    X  \\  
     140   & X &   X  &	X &    X  &  X  &    -  &    X  &    X  &    X  \\  
     141   & X &   X  &	X &    X  &  X  &    -  &    X  &    X  &    X  \\  
     142   & X &   X  &	X &    X  &  X  &    X  &    X  &    X  &    X  \\  
     143   & X &   X  &	X &    X  &  X  &    X  &    X  &    -  &    X  \\  
     144   & X &   X  &	X &    X  &  X  &    X  &    X  &    X  &    X  \\  
     145   & X &   X  &	X &    X  &  X  &    -  &    X  &    X  &    X  \\  
     146   & X &   X  &	X &    X  &  X  &    X  &    X  &    -  &    X  \\  
     147   & X &   X  &	X &    X  &  X  &    -  &    X  &    -  &    X  \\  
     148   & X &   X  &	X &    X  &  X  &    X  &    X  &    X  &    X  \\  
     149   & X &   X  &	X &    X  &  X  &    X  &    X  &    X  &    X  \\  
     150   & X &   X  &	X &    X  &  X  &    -  &    -  &    X  &    X  \\  
     151   & X &   X  &	X &    X  &  X  &    -  &    X  &    X  &    X  \\  
     152   & X &   X  &	X &    X  &  X  &    X  &    X  &    X  &    X  \\  
     153   & X &   X  &	X &    X  &  X  &    X  &    X  &    X  &    X  \\  
     154   & X &   X  &	X &    X  &  X  &    -  &    X  &    X  &    X  \\  
     155   & X &   X  &	X &    -  &  -  &    -  &    X  &    -  &    X  \\  
     156   & X &   X  &	X &    X  &  X  &    X  &    X  &    X  &    X  \\  
     157   & X &   X  &	X &    X  &  X  &    -  &    X  &    X  &    X  \\  
     158   & X &   X  &	X &    X  &  X  &    X  &    X  &    X  &    X  \\  
     159   & X &   X  &	X &    X  &  X  &    X  &    X  &    X  &    X  \\  
     160   & X &   X  &	X &    X  &  X  &    X  &    X  &    X  &    X  \\  
     161   & X &   X  &	X &    X  &  X  &    -  &    X  &    X  &    X  \\  
     162   & X &   X  &	X &    -  &  X  &    -  &    X  &    X  &    X  \\  
     163   & X &   X  &	X &    X  &  X  &    X  &    X  &    X  &    X  \\  
     164   & X &   X  &	X &    X  &  -  &    -  &    X  &    X  &    X  \\  
     165   & X &   X  &	X &    X  &  -  &    -  &    X  &    -  &    X  \\  
     166   & X &   X  &	X &    -  &  -  &    -  &    X  &    -  &    X  \\  
     167   & X &   X  &	X &    X  &  X  &    -  &    X  &    -  &    X  \\  
     168   & X &   X  &	X &    X  &  X  &    X  &    X  &    -  &    X  \\  
     169   & X &   X  &	X &    X  &  X  &    -  &    X  &    -  &    X  \\  
     170   & X &   X  &	X &    X  &  X  &    X  &    X  &    X  &    X  \\  
     171   & X &   X  &	X &    X  &  X  &    X  &    X  &    X  &    X  \\  
     172   & X &   X  &	X &    X  &  X  &    X  &    X  &    X  &    X  \\  
     173   & X &   X  &	X &    X  &  X  &    X  &    X  &    X  &    X  \\  
     174   & X &   X  &	X &    X  &  X  &    -  &    X  &    X  &    X  \\  
     175   & X &   X  &	X &    X  &  -  &    -  &    X  &    -  &    X  \\  
     176   & X &   X  &	X &    X  &  X  &    -  &    X  &    -  &    X  \\  
     177   & X &   X  &	X &    X  &  X  &    X  &    X  &    X  &    X  \\  
     178   & X &   X  &	X &    X  &  -  &    X  &    X  &    X  &    X  \\  
     179   & X &   X  &	X &    X  &  -  &    -  &    -  &    -  &    X  \\  
     180   & X &   X  &	X &    -  &  X  &    -  &    X  &    -  &    X  \\  
\noalign{\smallskip}
\hline
\end{tabular}
\]
\end{table*}

\addtocounter{table}{-1}
\begin{table*}
\tiny
\caption{continue}
\label{Tabmod}
\[\begin{tabular}{rccccccccc}
\hline
\noalign{\smallskip}
     HRS& 2MASS & SDSS & UV & H$\alpha$ & IRAS & 20cm & HI & CO & Spec \\
\noalign{\smallskip}    
\hline
\noalign{\smallskip}          
     181   & X &   X  &	X &    -  &  -  &    -  &    -  &    -  &    -  \\  
     182   & X &   X  &	X &    X  &  X  &    X  &    X  &    X  &    X  \\  
     183   & X &   X  &	X &    X  &  -  &    X  &    -  &    X  &    X  \\  
     184   & X &   X  &	X &    X  &  X  &    -  &    X  &    X  &    X  \\  
     185   & X &   X  &	X &    X  &  -  &    -  &    X  &    X  &    X  \\  
     186   & X &   X  &	- &    -  &  -  &    -  &    X  &    X  &    -  \\  
     187   & X &   X  &	X &    X  &  X  &    -  &    X  &    X  &    X  \\  
     188   & X &   X  &	X &    X  &  X  &    X  &    X  &    X  &    X  \\  
     189   & X &   X  &	X &    X  &  X  &    X  &    X  &    -  &    X  \\  
     190   & X &   X  &	X &    X  &  X  &    X  &    X  &    X  &    X  \\  
     191   & X &   X  &	X &    X  &  X  &    -  &    X  &    -  &    X  \\  
     192   & X &   X  &	X &    X  &  X  &    -  &    X  &    -  &    X  \\  
     193   & X &   X  &	X &    X  &  X  &    X  &    X  &    X  &    X  \\  
     194   & X &   X  &	X &    X  &  X  &    -  &    X  &    X  &    X  \\  
     195   & X &   X  &	- &    X  &  -  &    -  &    X  &    X  &    X  \\  
     196   & X &   X  &	X &    X  &  X  &    X  &    X  &    X  &    X  \\  
     197   & X &   X  &	X &    X  &  X  &    X  &    X  &    X  &    X  \\  
     198   & X &   X  &	X &    X  &  X  &    -  &    X  &    X  &    X  \\  
     199   & X &   X  &	X &    X  &  X  &    -  &    X  &    -  &    X  \\  
     200   & X &   X  &	X &    X  &  X  &    X  &    X  &    X  &    -  \\  
     201   & X &   X  &	X &    X  &  X  &    X  &    X  &    X  &    X  \\  
     202   & X &   X  &	- &    -  &  -  &    -  &    -  &    -  &    X  \\  
     203   & X &   X  &	X &    X  &  X  &    X  &    X  &    X  &    X  \\  
     204   & X &   X  &	X &    X  &  X  &    X  &    X  &    X  &    X  \\  
     205   & X &   X  &	X &    X  &  X  &    X  &    X  &    X  &    X  \\  
     206   & X &   X  &	X &    X  &  X  &    X  &    X  &    X  &    X  \\  
     207   & X &   X  &	X &    X  &  X  &    X  &    X  &    X  &    X  \\  
     208   & X &   X  &	X &    X  &  X  &    -  &    X  &    X  &    X  \\  
     209   & X &   -  &	X &    -  &  X  &    X  &    X  &    -  &    -  \\  
     210   & X &   X  &	X &    -  &  -  &    -  &    X  &    -  &    X  \\  
     211   & X &   X  &	X &    X  &  X  &    X  &    -  &    -  &    X  \\  
     212   & X &   X  &	X &    X  &  X  &    X  &    X  &    X  &    X  \\  
     213   & X &   X  &	X &    X  &  X  &    X  &    X  &    X  &    X  \\  
     214   & X &   X  &	X &    -  &  -  &    -  &    X  &    -  &    -  \\  
     215   & X &   X  &	X &    X  &  -  &    X  &    X  &    X  &    X  \\  
     216   & X &   X  &	X &    X  &  X  &    X  &    X  &    X  &    X  \\  
     217   & X &   X  &	X &    X  &  X  &    X  &    X  &    X  &    X  \\  
     218   & X &   X  &	X &    -  &  -  &    -  &    X  &    -  &    -  \\  
     219   & X &   X  &	- &    -  &  -  &    -  &    X  &    -  &    -  \\  
     220   & X &   X  &	X &    X  &  X  &    X  &    X  &    X  &    X  \\  
     221   & X &   X  &	X &    X  &  X  &    -  &    X  &    X  &    X  \\  
     222   & X &   X  &	X &    X  &  -  &    -  &    X  &    -  &    X  \\  
     223   & X &   X  &	X &    X  &  -  &    -  &    X  &    -  &    X  \\  
     224   & X &   X  &	X &    X  &  X  &    -  &    X  &    X  &    X  \\  
     225   & X &   X  &	X &    X  &  -  &    -  &    X  &    -  &    X  \\  
     226   & X &   X  &	X &    X  &  X  &    -  &    X  &    -  &    X  \\  
     227   & X &   X  &	X &    X  &  X  &    X  &    X  &    -  &    X  \\  
     228   & X &   -  &	- &    -  &  -  &    -  &    -  &    -  &    -  \\  
     229   & X &   X  &	X &    -  &  -  &    -  &    X  &    -  &    X  \\  
     230   & X &   X  &	X &    X  &  X  &    X  &    X  &    X  &    X  \\  
     231   & X &   X  &	- &    X  &  X  &    -  &    X  &    X  &    X  \\  
     232   & X &   X  &	X &    X  &  X  &    -  &    X  &    X  &    X  \\  
     233   & X &   X  &	X &    X  &  X  &    X  &    X  &    X  &    X  \\  
     234   & X &   X  &	- &    -  &  -  &    X  &    X  &    -  &    X  \\  
     235   & X &   X  &	X &    -  &  -  &    -  &    X  &    -  &    -  \\  
     236   & X &   X  &	X &    -  &  -  &    -  &    -  &    -  &    X  \\  
     237   & X &   X  &	X &    X  &  X  &    X  &    X  &    X  &    X  \\  
     238   & X &   X  &	- &    -  &  -  &    -  &    X  &    -  &    X  \\  
     239   & X &   X  &	X &    X  &  X  &    X  &    X  &    X  &    X  \\  
     240   & X &   X  &	X &    -  &  -  &    -  &    X  &    -  &    X  \\  
     241   & X &   X  &	X &    X  &  -  &    X  &    X  &    X  &    -  \\  
     242   & X &   X  &	X &    X  &  X  &    X  &    X  &    X  &    X  \\  
     243   & X &   X  &	X &    X  &  X  &    -  &    X  &    -  &    -  \\  
     244   & X &   X  &	X &    X  &  X  &    X  &    X  &    X  &    X  \\  
     245   & X &   X  &	X &    X  &  X  &    X  &    X  &    X  &    X  \\  
     246   & X &   X  &	X &    X  &  X  &    X  &    X  &    X  &    X  \\  
     247   & X &   X  &	X &    X  &  X  &    X  &    X  &    X  &    X  \\  
     248   & X &   X  &	X &    -  &  -  &    -  &    -  &    -  &    X  \\  
     249   & X &   X  &	X &    X  &  -  &    -  &    X  &    -  &    X  \\  
     250   & X &   X  &	X &    -  &  -  &    -  &    X  &    -  &    -  \\  
     251   & X &   X  &	X &    X  &  X  &    X  &    X  &    X  &    X  \\  
     252   & X &   X  &	X &    -  &  X  &    -  &    X  &    -  &    X  \\  
     253   & X &   X  &	X &    -  &  X  &    X  &    X  &    -  &    -  \\  
     254   & X &   X  &	- &    X  &  X  &    X  &    X  &    X  &    X  \\  
     255   & X &   X  &	X &    X  &  X  &    X  &    X  &    X  &    X  \\  
     256   & X &   X  &	X &    -  &  X  &    X  &    X  &    X  &    -  \\  
     257   & X &   X  &	X &    X  &  X  &    -  &    X  &    X  &    X  \\  
     258   & X &   X  &	X &    X  &  X  &    -  &    -  &    X  &    -  \\  
     259   & X &   X  &	- &    X  &  X  &    X  &    X  &    X  &    X  \\  
     260   & X &   X  &	X &    X  &  X  &    X  &    X  &    X  &    X  \\  
     261   & X &   X  &	X &    -  &  X  &    -  &    X  &    -  &    X  \\  
     262   & X &   X  &	X &    X  &  X  &    X  &    X  &    X  &    X  \\  
     263   & X &   X  &	X &    X  &  X  &    X  &    X  &    X  &    X  \\  
     264   & X &   X  &	X &    -  &  -  &    -  &    X  &    -  &    X  \\  
     265   & X &   -  &	X &    -  &  X  &    X  &    -  &    -  &    X  \\  
     266   & X &   -  &	X &    X  &  X  &    X  &    X  &    X  &    X  \\  
     267   & X &   X  &	X &    X  &  X  &    X  &    X  &    -  &    X  \\  
     268   & X &   X  &	X &    X  &  X  &    X  &    X  &    X  &    X  \\  
     269   & X &   X  &	X &    -  &  -  &    -  &    X  &    -  &    X  \\  
     270   & X &   X  &	X &    -  &  X  &    -  &    X  &    X  &    X  \\  
\noalign{\smallskip}
\hline
\end{tabular}
\]
\end{table*}

\addtocounter{table}{-1}
\begin{table*}
\tiny
\caption{continue}
\label{Tabmod}
\[\begin{tabular}{rccccccccc}
\hline
\noalign{\smallskip}
     HRS& 2MASS & SDSS & UV & H$\alpha$ & IRAS & 20cm & HI & CO & Spec \\
\noalign{\smallskip}    
\hline
\noalign{\smallskip}          
     271   & X &   X  &	X &    X  &  X  &    X  &    X  &    -  &    X  \\  
     272   & X &   X  &	X &    -  &  -  &    -  &    X  &    -  &    X  \\  
     273   & X &   X  &	X &    X  &  X  &    -  &    X  &    X  &    X  \\  
     274   & X &   X  &	X &    X  &  X  &    -  &    X  &    X  &    X  \\  
     275   & X &   -  &	X &    X  &  X  &    X  &    X  &    X  &    X  \\  
     276   & X &   X  &	X &    X  &  X  &    X  &    X  &    X  &    X  \\  
     277   & X &   X  &	X &    -  &  -  &    -  &    X  &    -  &    X  \\  
     278   & X &   X  &	- &    X  &  -  &    -  &    X  &    -  &    X  \\  
     279   & X &   X  &	X &    X  &  X  &    -  &    X  &    -  &    X  \\  
     280   & X &   X  &	X &    X  &  X  &    X  &    X  &    X  &    X  \\  
     281   & X &   X  &	X &    X  &  -  &    -  &    X  &    -  &    X  \\  
     282   & X &   X  &	- &    -  &  -  &    -  &    -  &    -  &    X  \\  
     283   & X &   X  &	X &    X  &  X  &    X  &    X  &    X  &    X  \\  
     284   & X &   -  &	X &    -  &  X  &    X  &    X  &    X  &    X  \\  
     285   & X &   X  &	X &    X  &  X  &    X  &    X  &    X  &    X  \\  
     286   & X &   X  &	X &    X  &  -  &    -  &    X  &    X  &    X  \\  
     287   & X &   X  &	X &    X  &  X  &    X  &    X  &    X  &    X  \\  
     288   & X &   -  &	X &    X  &  X  &    X  &    X  &    X  &    X  \\  
     289   & X &   -  &	X &    X  &  X  &    X  &    X  &    X  &    X  \\  
     290   & X &   X  &	X &    X  &  X  &    X  &    X  &    -  &    X  \\  
     291   & X &   X  &	X &    -  &  -  &    -  &    X  &    X  &    -  \\  
     292   & X &   X  &	X &    X  &  X  &    X  &    X  &    X  &    X  \\  
     293   & X &   X  &	X &    X  &  X  &    X  &    X  &    X  &    X  \\  
     294   & X &   X  &	- &    X  &  X  &    X  &    X  &    -  &    X  \\  
     295   & X &   X  &	X &    X  &  X  &    X  &    X  &    X  &    X  \\  
     296   & X &   X  &	X &    X  &  X  &    X  &    X  &    -  &    -  \\  
     297   & X &   X  &	X &    X  &  X  &    X  &    X  &    X  &    X  \\  
     298   & X &   X  &	X &    X  &  X  &    X  &    X  &    -  &    X  \\  
     299   & X &   X  &	X &    X  &  X  &    -  &    X  &    X  &    X  \\  
     300   & X &   X  &	- &    X  &  X  &    -  &    X  &    -  &    X  \\  
     301   & X &   X  &	X &    X  &  X  &    -  &    X  &    X  &    X  \\  
     302   & X &   X  &	X &    X  &  X  &    -  &    X  &    -  &    X  \\  
     303   & X &   X  &	X &    X  &  X  &    X  &    -  &    -  &    X  \\  
     304   & X &   X  &	X &    X  &  X  &    -  &    X  &    X  &    X  \\  
     305   & X &   X  &	X &    X  &  -  &    -  &    X  &    -  &    X  \\  
     306   & X &   X  &	X &    -  &  X  &    X  &    X  &    X  &    X  \\  
     307   & X &   X  &	X &    X  &  X  &    X  &    X  &    X  &    X  \\  
     308   & X &   X  &	X &    X  &  -  &    -  &    -  &    -  &    X  \\  
     309   & X &   X  &	- &    X  &  X  &    -  &    X  &    -  &    X  \\  
     310   & X &   X  &	X &    X  &  -  &    X  &    X  &    X  &    X  \\  
     311   & X &   X  &	X &    X  &  X  &    X  &    X  &    X  &    X  \\  
     312   & X &   X  &	X &    -  &  X  &    -  &    X  &    -  &    -  \\  
     313   & X &   X  &	X &    X  &  X  &    X  &    X  &    X  &    X  \\  
     314   & X &   X  &	X &    X  &  X  &    X  &    X  &    -  &    X  \\  
     315   & X &   X  &	X &    X  &  -  &    -  &    X  &    -  &    X  \\  
     316   & X &   X  &	X &    -  &  -  &    -  &    X  &    -  &    -  \\  
     317   & X &   X  &	X &    -  &  -  &    -  &    X  &    -  &    X  \\  
     318   & X &   X  &	X &    X  &  X  &    X  &    X  &    X  &    X  \\  
     319   & X &   X  &	X &    X  &  X  &    X  &    X  &    X  &    X  \\  
     320   & X &   X  &	X &    X  &  X  &    X  &    X  &    X  &    X  \\  
     321   & X &   X  &	X &    -  &  X  &    X  &    X  &    -  &    X  \\  
     322   & X &   X  &	X &    X  &  X  &    -  &    X  &    -  &    X  \\  
     323   & X &   X  &	X &    X  &  X  &    X  &    X  &    X  &    X  \\  
\noalign{\smallskip}
\hline
\end{tabular}
\]
\end{table*}

\clearpage

\end{document}